\documentclass[twocolumn]{aastex631}

\usepackage{xspace}
\usepackage{color}

\DeclareRobustCommand{\ion}[2]{%
\relax\ifmmode
\ifx\testbx\f@series
{\mathbf{#1\,\mathsc{#2}}}\else
{\mathrm{#1\,\mathsc{#2}}}\fi
\else\textup{#1\,{\mdseries\textsc{#2}}}%
\fi}

\def\arcsec{\hbox{$^{\prime\prime}$}\xspace}

\newcommand{\Msunyr}{\hbox{$M_\odot \,\hbox{yr}^{-1}$}\xspace}

\newcommand{\rs}{$R_{\odot}$\xspace}
\newcommand{\ls}{$L_{\odot}$\xspace}
\newcommand{\kms}{km\,s$^{-1}$\xspace}

\newcommand{\ec}{$\eta$\,Car\xspace}

\newcommand{\ha}{H$\alpha$\xspace}
\newcommand{\hd}{H$\delta$\xspace}

\newcommand{\hg}{H$\gamma$\xspace}

\newcommand{\red}{\color{red}}

\DeclareRobustCommand{\ion}[2]{%
\relax\ifmmode
\ifx\testbx\f@series
{\mathbf{#1\,\mathsc{#2}}}\else
{\mathrm{#1\,\mathsc{#2}}}\fi
\else\textup{#1\,{\mdseries\textsc{#2}}}%
\fi}

\begin{document}

\title[pcygabs-midcycle]{ 
 The long-term spectral changes of eta Carinae: are they caused by a dissipating occulter as indicated by {\sc cmfgen} models?} 

\correspondingauthor{Damineli, Augusto}
\email{augusto.damineli@iag.usp.br}

\author[0000-0002-7978-2994]{Damineli,~Augusto}
\affiliation{Universidade de S\~ao Paulo, Instituto de Astronomia, Geof\'isica e Ci\^encias Atmosf\'ericas, \\ Rua do Mat\~ao 1226, Cidade Universit\'aria, S\~ao Paulo, Brasil}

\author[0000-0001-5094-8017]{Hillier,~Desmond J.}
\affiliation{Department of Physics and Astronomy{FE}{ii} \& Pittsburgh Particle Physics, Astrophysics, and Cosmology Center (PITT PACC), \\ University of Pittsburgh, 3941 O'Hara Street,  Pittsburgh, PA 15260, USA}

\author[0000-0002-0284-0578]{Navarete,~Felipe}
\affiliation{SOAR Telescope/NSF's NOIRLab, Avda Juan Cisternas 1500, 1700000, La Serena, Chile}

\author[0000-0002-4333-9755]{Moffat,~Anthony ~F.~J}
\affiliation{D\'epartement de Physique and Centre de Recherche en Astrophysique du Qu\'ebec (CRAQ)\\ Universit\'e de Montr\'eal, C.P. 6128, Succ. Centre-Ville, Montr\'eal, Qu\'ebec, H3C 3J7, Canada}

\author[0000-0001-9754-2233]{Weigelt,~Gerd}
\affiliation{Max Planck Institute for Radio Astronomy, Auf dem H\"{u}gel 69, D-53121 Bonn, Germany}

\author[0000-0002-7762-3172]{Corcoran,~Michael~F.}
\affiliation{CRESST II \& X-ray Astrophysics Laboratory, NASA/Goddard Space Flight Center, Greenbelt, MD 20771, USA}
\affiliation{The Catholic University of America, 620 Michigan Avenue N.E., Washington, DC 20064, USA}

\author[0000-0002-6851-5380]{Gull,~Theodore.~R.}
\affiliation{Exoplanets \& Stellar Astrophysics Laboratory, NASA/Goddard Space Flight Center, Greenbelt, MD 20771, USA}

\author[0000-0002-2806-9339]{Richardson,~Noel D.}
\affiliation{Department of Physics and Astronomy, Embry-Riddle Aeronautical University, 3700 Willow Creek Road, Prescott, AZ 86301, USA}

\author{Ho,~Peter}
\affiliation{Department of Applied Mathematics, University of California, Santa Cruz, 1156 High Street, Santa Cruz, CA 95064, USA}

\author[0000-0001-7697-2955]{Madura,~Thomas~I.}
\affiliation{Department of Physics and Astronomy, San Jos\'e State University, One Washington Square, San Jos\'e, CA 95192-0106, USA}

\author[0000-0003-2971-0439]{Espinoza-Galeas,~David}
\affiliation{Departamento de Astronomia y Astrofisica, Facultad de Ciencias Espaciales, Universidad Nacional Autonoma de Honduras, Bulevar Suyapa, Tegucigalp{FE}{ii}a, M.D.C, Honduras, Centroamerica}

\author[0000-0001-9853-2555]{Hartman,~Henrik}
\affiliation{Materials Science and Applied Mathematics, Malm\"o University, SE-20506 Malm\"o, Sweden}

\author[0000-0002-5186-4381]{Morris,~Patrick}
\affiliation{California Institute of Technology, IPAC, M/C 100-22, Pasadena, CA 91125, USA}

\author{Pickett,~Connor~S.}
\affiliation{Department of Physics and Astronomy, Embry-Riddle Aeronautical University, 3700 Willow Creek Road, Prescott, AZ 86301, USA}

\author[0000-0001-7673-4340]{Stevens,~Ian~R.}
\affiliation{School of Physics \& Astronomy, University of Birmingham, Birmingham B15 2TT, UK}

\author[0000-0002-1942-7296]{Russell,~Christopher~M.~P.}
\affiliation{Department of Physics and Astronomy, University of Delaware, Newark, DE 19716, USA}

\author[0000-0001-7515-2779]{Hamaguchi,~Kenji}
\affiliation{CRESST II and X-ray Astrophysics Laboratory, NASA/Goddard Space Flight Center, Greenbelt, MD 20771, USA}
\affiliation{Department of Physics, University of Maryland Baltimore County, 1000 Hilltop Circle, Baltimore, MD 21250, USA}

\author[0000-0002-0386-2306]{Jablonski,~Francisco~J.}
\affiliation{Instituto Nacional de Pesquisas Espaciais/MCTIC Avenida dos Astronautas 1758, S\~ao Jos\'e dos Campos, SP, 12227-010, Brazil}

\author[0000-0002-8289-3660]{Teodoro,~Mairan}
\affiliation{Space Telescope Science Institute, 3700 San Martin Dr, Baltimore, MD 21218}

\author{McGee,~Padric}
\affiliation{Department of Physics, School of Physical Sciences, University of Adelaide, South Australia, 5005}
\affiliation{SASER Team, 269 Domain Road, South Yarra, Vic 3141, Australia}

\author{Cacella,~Paulo}
\affiliation{SASER Team, 269 Domain Road, South Yarra, Vic 3141, Australia}

\author{Heathcote,~Bernard}
\affiliation{SASER Team, 269 Domain Road, South Yarra, Vic 3141, Australia}

\author{Harrison,~Ken~M.}
\affiliation{SASER Team, 269 Domain Road, South Yarra, Vic 3141, Australia}

\author{Johnston,~Mark}
\affiliation{SASER Team, 269 Domain Road, South Yarra, Vic 3141, Australia}

\author{Bohlsen,~Terry}
\affiliation{SASER Team, 269 Domain Road, South Yarra, Vic 3141, Australia}

\author{Di Scala,~Giorgio}
\affiliation{SASER Team, 269 Domain Road, South Yarra, Vic 3141, Australia}

\begin{abstract}
{ Eta Carinae (\ec) exhibits a unique set of P Cygni profiles with both broad and narrow components. Over many decades, the spectrum has changed -- there has been an increase in observed continuum fluxes and a decrease in  \ion{Fe}{ii} and \ion{H}{i} emission line equivalent widths. The spectrum is evolving towards that of a P Cygni star such as P~Cygni itself and HDE~316285.  
The spectral evolution has been attributed to intrinsic variations such as a decrease in the mass-loss rate of the primary star or differential evolution in a latitudinal-dependent stellar wind. However intrinsic wind changes conflict with three observational results: the steady long-term bolometric luminosity; the repeating X-ray light curve over the binary period; and the constancy of the dust-scattered spectrum from the Homunculus. We extend previous work that showed a secular strengthening of P~Cygni absorptions by adding more orbital cycles to overcome temporary instabilities and by examining more atomic transitions. {\sc cmfgen} modeling of the primary wind shows that a time-decreasing  mass-loss rate is not the best explanation for the observations. However, models with a `small' dissipating absorber in our line-of-site can  explain both the increase in brightness and changes in the emission and P Cygni absorption profiles. If the spectral evolution is caused by the dissipating circumstellar medium, and not by intrinsic changes in the binary, the dynamical timescale to recover from the Great Eruption is much less than a century, different from previous suggestions.} 
\end{abstract}


\section{Introduction}
\label{sectionintroduction}

The \ec spectrum recorded in the 1980s was very unusual for an LBV (or P~Cygni) star \citep{hillier2001,humphreys94}. It exhibited broad wind lines (FWHM$ \sim 800$\,\kms) of \ion{H}{i}, \ion{Fe}{ii}, and \ion{He}{i}\, which are now understood to be associated with the primary stellar wind, although \ion{He}{i}\ is also likely to be influenced by the wind-wind collision (WWC) zone and the radiation of the secondary star. Very prominent narrow emission lines (FWHM$\sim$\,20--80\,\kms) of \ion{H}{i}, \ion{He}{i}, \ion{Fe}{ii} and [\ion{Fe}{ii}], superimposed on the spectrum, arise in the Weigelt clumps \citep{davidson95}. Both the \ion{Fe}{ii} and [\ion{Fe}{ii}] lines exhibited a broad base with similar FWHM values to that of the primary wind lines. The broad [\ion{Fe}{ii}] lines are unusual and are not seen in other stars with strong stellar winds at similar excitation regimes \citep[e.g. HDE 316285,][]{hillier2001}. They are now understood to originate in the stellar wind, but their strength relative to the stellar continuum has been enhanced by the occulter \citep{HDI01_eta_wind}. { The same effect makes the narrow components unexpectedly prominent.}

{ \citet{mehner19}, using the dusty Homunculus as a calorimeter in the infrared, demonstrated that the binary luminosity has been reasonably constant over the period 1968 to 2018 with possible variations associated with the 5.53-year binary orbit. Yet from 2000 to 2020 the stellar brightness increased by more than a decade in the UV \citep{Gull2021}.
The dust in the Homunculus absorbs the bulk of the  radiation  from the visible into the FUV with subsequent re-emission in the infrared. The pronounced stability of the X-ray light curve over the past five binary cycles \citep{espinoza21,corcoran17} confirms that the stellar winds and their collision have been stable over the past two decades.

\citet{mehner12} reported that the stellar continuum at 3950\AA\ reflected by Weigelt clump D stayed constant over two cycles (1999 to 2010) while the flux of the central star increased more than ten-fold (their Fig. 12).}

{ 
\citet{damineli19} suggested that a dissipating dusty structure that partially occulted the central star and the extended winds would explain the reports by \citet{gull09} and by \citet{Mehner2014} that while the central star brightened in the visible and UV, the Weigelt clumps \citep{weigelt86} remained at constant flux, leading to an effective decrease in line equivalent widths since the EW is the line flux divided by the continuum inside the slit.}

{
Long-term strength decreases of some broad emission lines have been reported based on both ground-based and STIS observations made by \cite{Mehner2010,mehner11,mehner12,mehner15,mehner19,martin21,davidson18}. Those authors suggested that the strength decrease might be due to the intrinsic evolution of the central star and that the secular brightness increase could be produced by a decrease in \ec's primary wind density or a decrease in circumstellar extinction. { Another suggestion was that the latitude-dependent wind evolves differently in the equator and polar regions \citep{mehner15}. A latitude-dependent 
stellar wind was previously proposed by \citet{smith03}, who invoked rotation of the star as the cause.  \citet{groh12} showed that these observations are more likely explained by the ionization impact of the hot secondary modifying the primary wind structure. Moreover, the fast-rotator LBVs show S-Doradus instabilities, not seen in slow rotators and not present in \ec, suggesting that \ec might be a slow rotator, although this is not what is expected in a binary star merger.}

{ The trend in the system's photometric brightness \citep[][]{damineli19} along with a study of direct versus reflected line emission \citep[][see also \citealt{mehner15}]{damineli21} -- suggested that the long-term weakening of the emission lines could be better explained by a drop in circumstellar extinction along our line-of-sight (LOS) without recourse to intrinsic changes in \ec~A.}

{
 The impact of the dissipating occulter on spectral features is complicated, since both the line
 and continuum forming regions in \ec are extended, and since different spectral lines are formed
 at different locations in the stellar wind. Thus, we first present a qualitative summary of the impact
 of an occulter on the intrinsice spectrum, which is complemented by the semi-quantitative {\sc cmfgen} model described in Sect.~\ref{section_cmfg}.}

Fig.~\ref{occulter} shows a schematic view of the five main regions contributing to the observed spectrum of \ec. Taking advantage of STIS/HST observations makes it possible to disentangle the spectrum of the Weigelt clumps (region C) with their nebular emission lines and reflected continuum  (which are relatively small in amplitude) from the central sources. { Typical seeing from the ground of 1 to 2 arcsec (2300 to 4600 au) while HST/STIS spatial resolution is 0\farcs1 (230 au)}. The spectrum reflected from the Homunculus (region D) can also be resolved (for example, at the FOS\,4 position) even from ground-based observations \citep{davidson97}. The central region is unresolved and is composed of $a)$ the central star plus its inner wind (region A) -- the major source of light in the system, the source of both the stellar continuum and the very broad P~Cygni spectral features profiles; and $b)$ the outer wind (region B) which contributes with relatively broad lines and some continuum flux. The occulter (region E) partially blocks the wind region but not the Weigelt clumps. 

{ The occulter apparently blocks the central stellar core and part of the extended primary wind. As it dissipates, the contribution from the primary wind increases by a factor $\sim$2 while the core (continuum) increased $\sim$10-fold.}

\begin{figure*}[ht!]
\centering
{\includegraphics[width=1\linewidth,angle=0,viewport=45bp 55bp 950bp 490bp, clip]{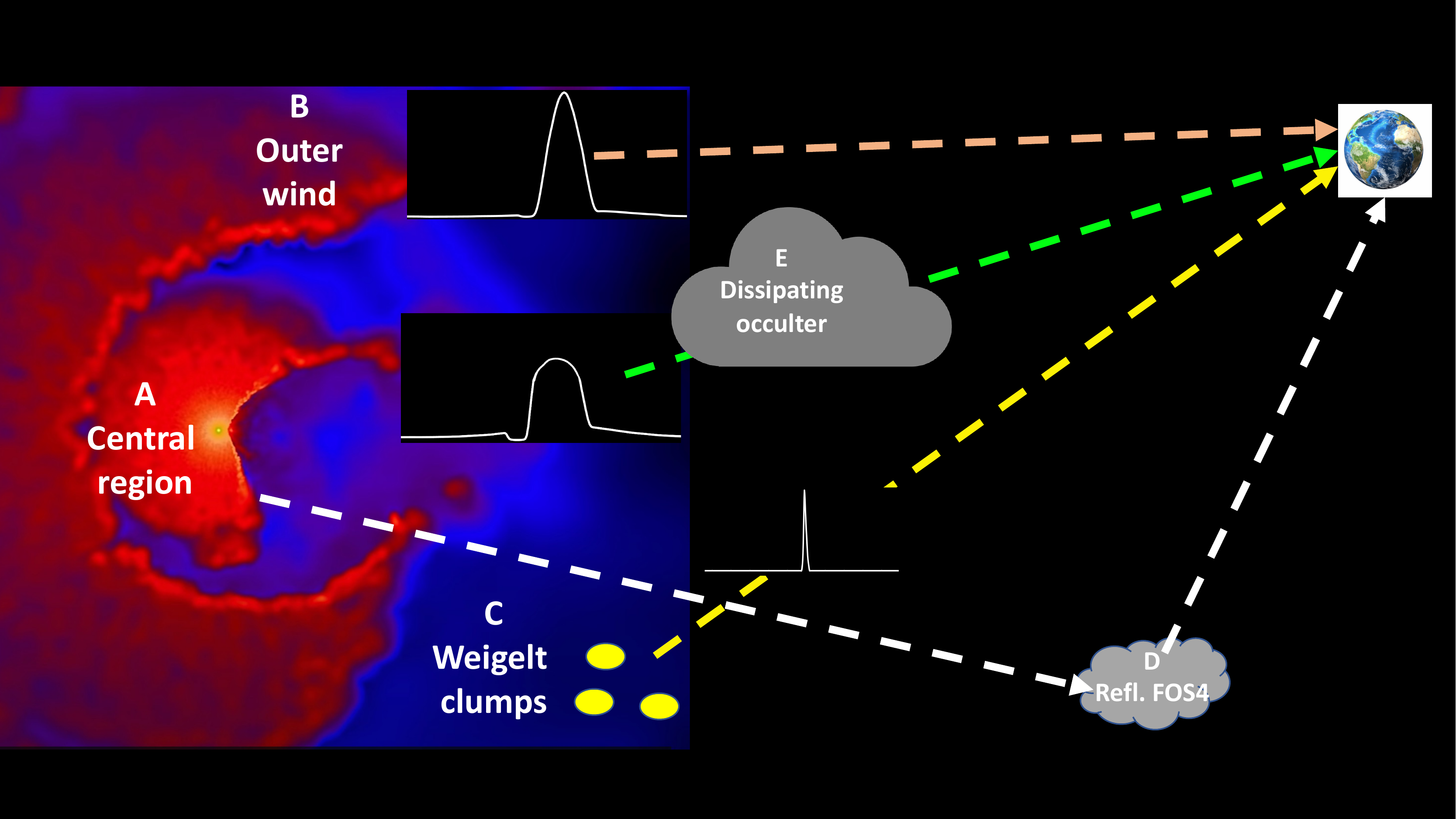}}
\caption{{ Scheme showing the five main components of the \ec spectrum.  The \ec image is a snapshot from a 3D simulation by \citet{madura13}. The structures are not to scale.}}
\label{occulter}
\end{figure*}

The FOS\,4 region at the SE pole of the Homunculus is located at {4\arcsec} from the central stellar core \citep{davidson05}. This region receives mostly unobstructed light from the central object (see the darker line in the upper plot of Fig.~\ref{4500_dirXrefl}), with a spectrum that is typical for an LBV star, with no significant contribution from narrow-line components. Its spectrum was taken with the UVES spectrograph in 2006.5 (seeing-limited, ground-based observations) and has remained very stable since then (see Fig.~4 from \citealt{damineli21})\footnote{Some intrinsic, low-excitation, line emission (such as [\ion{Ni}{ii}] $\lambda 7378$) arises in the Homunculus but exhibits a distinct velocity shift from the scattered emission \citep{hillier92}}. A coeval spectrum was also taken with UVES in direct view, indicated as the red spectrum in Fig.~\ref{4500_dirXrefl}.
{ The year 2006.5 corresponds to the mid-cycle of the \ec system ($\phi\,=\,0.52$) and fairly represents its high-excitation phase. The wavelength range was chosen to show lines from representative ionic species.}
The direct view of the star is attenuated by the (semi-transparent) occulter, which diminishes the flux originating from the central region, increasing the strength of narrow lines relative to the stellar continuum, from the perspective of the observer. { \citet{gull23} suggest that the occulter is located at $\approx$\,1000\,au from the central star and was ejected around 1890.} 

\begin{figure*}[ht!]
\centering
{\includegraphics[width=1\linewidth,angle=0,viewport=60bp 15bp 880bp 520bp, clip]{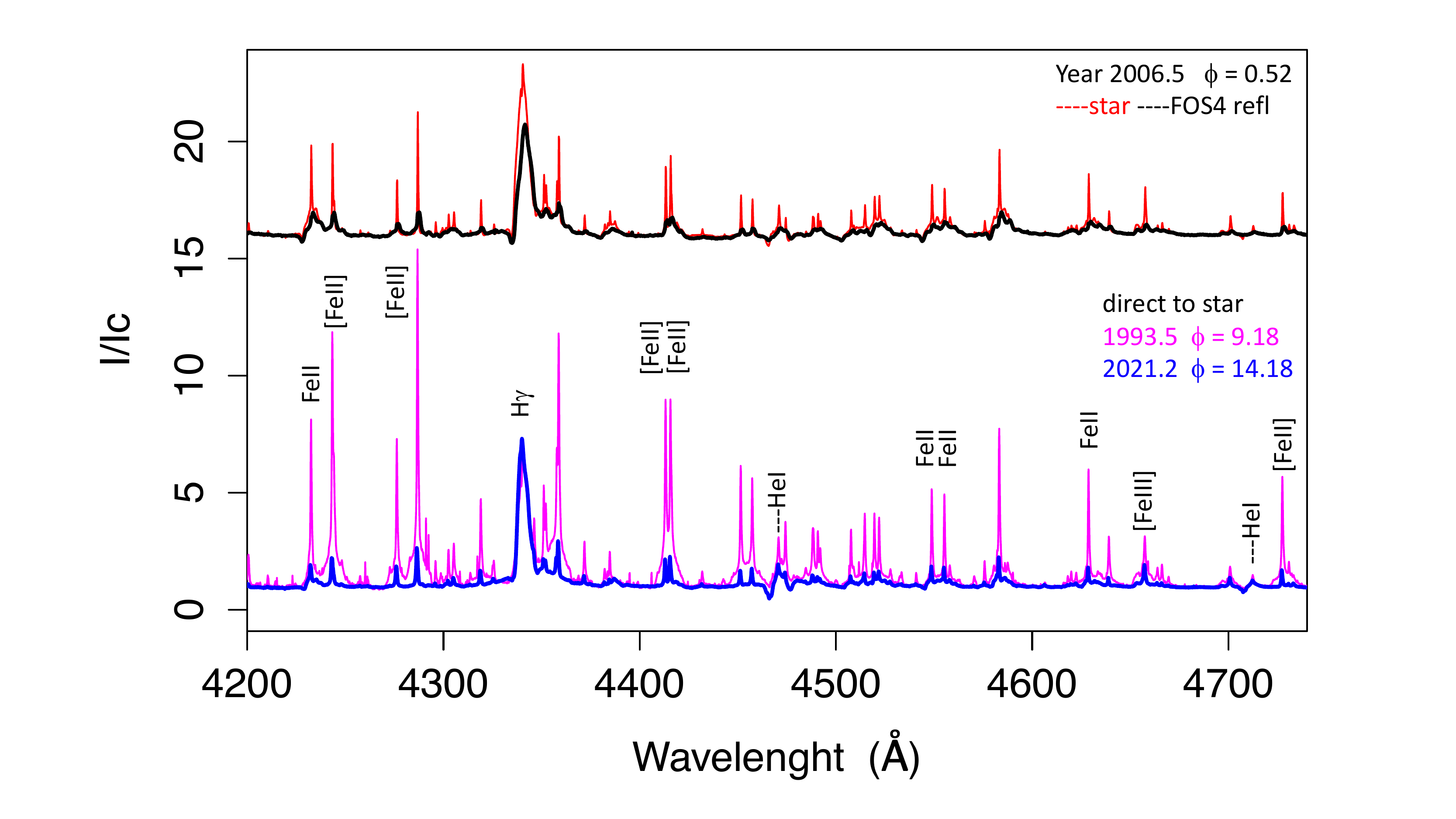}}
\caption{{ Ground-based (seeing-limited) high-resolution spectra of \ec}. {\it Upper panel:} The thick black line at the top shows the spectrum reflected off the Homunculus at the FOS\,4 position, observed in 2006.5. It represents an unobstructed view of the system. \ha emission profile in reflected light has been decreasing by a much smaller amount than in direct light and other emission lines have not changed. The thinner red line shows a coeval spectrum in direct view toward the central star (presumably impacted by the occulter).
{\it Bottom panel:} direct view spectra taken in 1993.5 ({ orbital phase 0.18) with the FEROS spectrograph (thin pink line) and 2021.2 (same phase), taken with LCOGT/NRES (thick blue line).  A broad base on many of the \ion{Fe}{ii} lines is readily apparent. The direct spectrum has evolved towards that reflected at FOS\,4, which has evolved by a small amount.}
}
\label{4500_dirXrefl}
\end{figure*}

{ As the occulter dissipates with time, the spectrum becomes more similar to that of the reflected view, as shown by comparing the pink line in the bottom plot of Fig.~\ref{4500_dirXrefl}, taken in 1993.5 using ESO/FEROS, with the overplotted blue spectrum taken in 2021.2 with LCOGT/NRES.)}

The effect of the occulter on the line strength is strongly dependent on the radial density profile of the line-formation zone plus the size, geometry, and radial opacity profile of the occulter. An approximation of the stellar wind structure can be {inferred} from the model of {\citet[][see their Fig.~9]{hillier2001}}. Broad \ion{Fe}{ii} lines are formed in the outer regions of the primary's wind. Their intensities relative to the continuum have been decreasing with time at a slower pace than the narrow lines formed in the Weigelt clumps. The broad {forbidden} [\ion{Fe}{ii}] lines are formed at typically larger radii than the {permitted} \ion{Fe}{ii} lines. In addition, their strength (normalized to the continuum) has also decreased by a {significant} factor from 1993.5 to 2021.2, as indicated in the bottom panel of Fig.~\ref{4500_dirXrefl}. During the same period (1993.5--2021.2), the narrow-line fluxes (normalized line strength times the continuum flux) remained constant \citep{damineli21}.

{ Higher energy excitation lines (e.g., \ion{He}{i}, \ion{N}{ii} and high members of the \ion{H}{i} line series) are formed close to the central star, and some of them are entirely covered by the occulter. For instance, the H\,${\gamma}$ line shown in Fig.~\ref{4500_dirXrefl} has not significantly changed over the last 28 years, while the EW of H$\alpha$ has decreased by a factor of $\sim$2. High-excitation permitted lines formed behind the occulter would maintain their normalized intensity if there was no emission from outside the occulter. However, the wind region emits both continuum (bound-free and free-free radiation, and electron scattering) plus line emission, and this can veil absorption lines. Further, if the occulter extinction is large enough, the observed spectrum will be dominated by the spectral formation in the region outside the occulter. As the flux from the occulted region increases (when the extinction decreases), the contribution from the central region increasingly dominates, making the absorption lines more visible.}

The presence of the hot secondary star complicates this simple scenario, as it affects both the primary wind structure and its ionization \citep[e.g.][]{mehner12,mehner11}. For most of the orbit, the higher-order Balmer lines do not exhibit a (strong) P~Cygni absorption component, which is directly attributable to the influence of the secondary \citep{groh12b}. However, strong  P~Cygni absorption just after periastron {(i.e., when the companion is behind the secondary and the trailing arm of the wind-wind collision structure is in the line-of-sight)} have been { reported by \citet{Nielsen07} and \citet{mehner15}.}

{ The goals of this work are $a)$ to report the long-term evolution of the P~Cygni absorption components as seen from ground-based spectroscopy; $b)$ to discuss how the strengthening of the absorption components and the weakening of the emission line components compare with models of decreasing mass-loss rate and with a dissipating occulter in front of \ec.}

This paper is organized as follows. 
In Sect.~\ref{sectionobservations}, we present the sources and general characteristics of our data; in Sect.~\ref{sectionresults}, we present the measurements and their analysis; in Sect.~\ref{section_cmfg}, we report {\sc cmfgen} modeling with normal and enhanced mass-loss rate and two versions of occulter. Sect.~\ref{sectiondiscussions}, we discuss our findings. 

\section{Data: observations and reduction}
\label{sectionobservations}
\begin{table}
	\centering									
	\caption{Observatories}
	\label{observatories}
	\begin{tabular}{lllllll} 
    \hline							
Observ.	& telesc. & Res.&5876&6347&8750&10830	\\
 & aper.& pow. &\#&\#&\#&\# \\
\hline
UVES	& 8m& 90K&1&62&28&-\\
GMOS& 8m& 4.4K&22&22&-&-\\
FEROS	& 1.5\&2.2m& 48K&186&169&141&- \\
OPD& 1.6m&22K&25&73&32&140\\
Hexapod	& 1.5m& 48K&8&-&-&- \\
CHIRON	& 1.5m& 90K&80&246&-&-\\
SMARTS	& 1.5m& 40K&91&83&-&-\\
NRES	& 1m& 48K&102&102&102&-\\
HERCULES& 1m& 48K&9&14&8&-\\
M.~Johnston& 0.60m&17K&-&5&-&-\\
P. mcGee& 0.35m&11K&1&7&-&-\\
P.~Cacella& 0.30m& 5.5K&13&-&-&-\\
K.~Harrison& 0.28m&10K&-&5&-&-\\
B.~Heatcote &0.28m &16K&10&26&-\\
T.~Bohlsen& 0.27m& 15K&3&12&-&- \\
G.~Di~Scala& 0.20m& 17K&-&2&-&-\\
STIS& 2.5m& 10K&-&-&-&- \\
\hline
\end{tabular}
Column 1 is the spectrograph/observer name; col. 2 the telescope aperture in meters; col. 3 the most frequent resolving power;  col. 4--7 the number of measurements used in this work.
\end{table}

{ For the modeling, and considering we are interested in looking at the general behavior of line strengths, it is sufficient to have two or more spectra taken at the exact same orbital phase over a time interval long enough to enhance the emission weakening and absorption strengthening effects. For the emission lines, it is already proven that there is a weakening trend in the long-term \citep{mehner15, damineli21}. For the P~Cygni absorptions, however, the past reports are confined to at most three cycles \citep{mehner15} and are better sampled around periastron. The fluctuations are large even at mid-cycle, so the idea that they are strengthening over many cycles is not completely stabilished. To show this, we require dense monitoring over as many cycles as possible. The most practical quantity to use for such a time series is the equivalent width (EW). }

The data used to measure EWs in this work are {mainly} from the facilities described in \citet{damineli21}. We list those facilities in Table~\ref{observatories} and, for details, we direct the reader to Sect.~2 of \citet{damineli21}. In addition to the professional facilities, there are a few observations from amateur astronomers -- these also listed in Table~\ref{observatories} and partly described in \citet{teodoro16} (also see the Acknowledgments).

Most of the spectra obtained from other authors and observatories were in 1D format (in counts versus $\lambda$), { reduced with standard procedures. They are seeing-limited ($>$\,1"), and since most of the signal from the object originates in the central arcsec, these spectra sample similar nebular contamination}. Data from ESO/UVES and Gemini/GMOS, downloaded as 2D images from the Treasury Project\footnote{\url{http://etacar.umn.edu/}}, were used to extract 1D spectra using all the spatial rows along the slit containing the stellar signal { to mimic the worse image quality of the other observations which used a slit (1" wide) or fiber ($\sim$3" diameter). Even in the case of excellent seeing, when some contribution to the signal from the Weigelt clumps could be left outside the slit aperture, they impact only the narrow-line components, and this work is concerned with the broad components, originating in a smaller region. 
The two HST/STIS 2D spectra were downloaded from  MAST site\footnote{\url{https://mast.stsci.edu/}}. To avoid as much as possible contamination by the Weigelt clumps in STIS spectro-images, we co-added five rows around the peak of the stellar image (0\farcs25-wide). Typical signal-to-noise ratios (S$/$N) of the ground-based spectra range between 60 and 300 in the central region, with the most frequent value $\sim$150. On average, the lowest S$/$N was observed for the P~Cygni absorption component of \ion{He}{i}~$\lambda$10830 line, with S/N$\sim$\,80 to 100 per pixel. This is because we use normal CCDs with very low quantum efficiency at those wavelengths ($\sim$\,1-2\%) and a large impact because of fringing and telluric lines. Despite \ec being a very bright source at that wavelength, observation of a telluric standard star is required. Other time-consuming procedures are applied in the flat-fielding to maximize the final S/N, demanding a few hours to get a reliable spectrum.}   

{ Spectra spanning a wide wavelength range were normalized to unity by dividing them by a high-order polynomial fit through line-free spectral windows representing the stellar continuum. This procedure was mostly used for visualization purposes like that presented in Fig.~\ref{4500_dirXrefl}. A linear fit was used for estimating the continuum by adopting line-free regions for spectral ranges shorter than $\sim$\,100\AA. 

{
For EWs of absorption components, we fitted a straight line representing the continuum across each spectral line. Two windows were chosen, one at each side of the line, spanning sixteen pixels. For \ion{He}{i}~$\lambda$5875 the continuum windows are centered at $\lambda$5851 and $\lambda$5930; $\lambda$6329 and $\lambda$6400 for \ion{Si}{ii}~$\lambda$6347; $\lambda$8722 and $\lambda$8770 for \ion{H}{i}~$\lambda$8752; $\lambda$10760 and $\lambda$11000 for \ion{He}{i}~$\lambda$10830. The S$/$N ratio of the continuum was measured inside the two continuum windows. The real S$/$N of the EW is lower for the absorption and higher for the emission, because were measured in the continuum windows.}

EWs were measured by direct integration in the area between that line profile and the projected stellar continuum. The uncertainty of individual EW measurements ($\sigma_{\rm EW}$) was evaluated using Eq.\,(3) from \citet{vollmann06}, given by:
\begin{equation}
    \sigma_{\rm EW} = \sqrt{2} \cdot \left( \frac{\Delta\lambda - \rm{EW}}{\rm{S/N}}\right)
    \label{eq_ewerror}
\end{equation}
\noindent where $\Delta$$\lambda$ is the wavelength interval used to integrate the flux (in \AA). Eq.\,\ref{eq_ewerror} {also accounts for the influence of} small-scale irregularities and possible faint lines in the continuum window. EW measurements with respective errors are reported in Appendix~\ref{onlinetables}, and presented in Fig.~\ref{timeseries}. 

{ For our high-quality spectra, the largest uncertainties might not be due to photon statistics, but rather arise from the procedures of continuum flattening and the bias introduced by the nebular contribution which will
not be identical for each spectrograph. The only way to evaluate these uncertainties is by comparing the coeval series of measurements taken with different spectrographs. We did this for the epochs for which it was possible compare measurements of at least two spectrographs. Only the EWs from the line \ion{Si}{i}~$\lambda$6437 had overlaps between different spectrographs that were frequent enough for this procedure. Unfortunately, the EW curve for this line is very complicated, with large variations within a couple of months. The scatter inside each data-sample was larger than the differences between its average curve and the other group used to compare with it. So, no relative shifts were applied for any EW groups.} 
\begin{figure*}[ht!]
\centering
{\includegraphics[width=\linewidth,viewport=0bp 25bp 1000bp 535bp, clip]{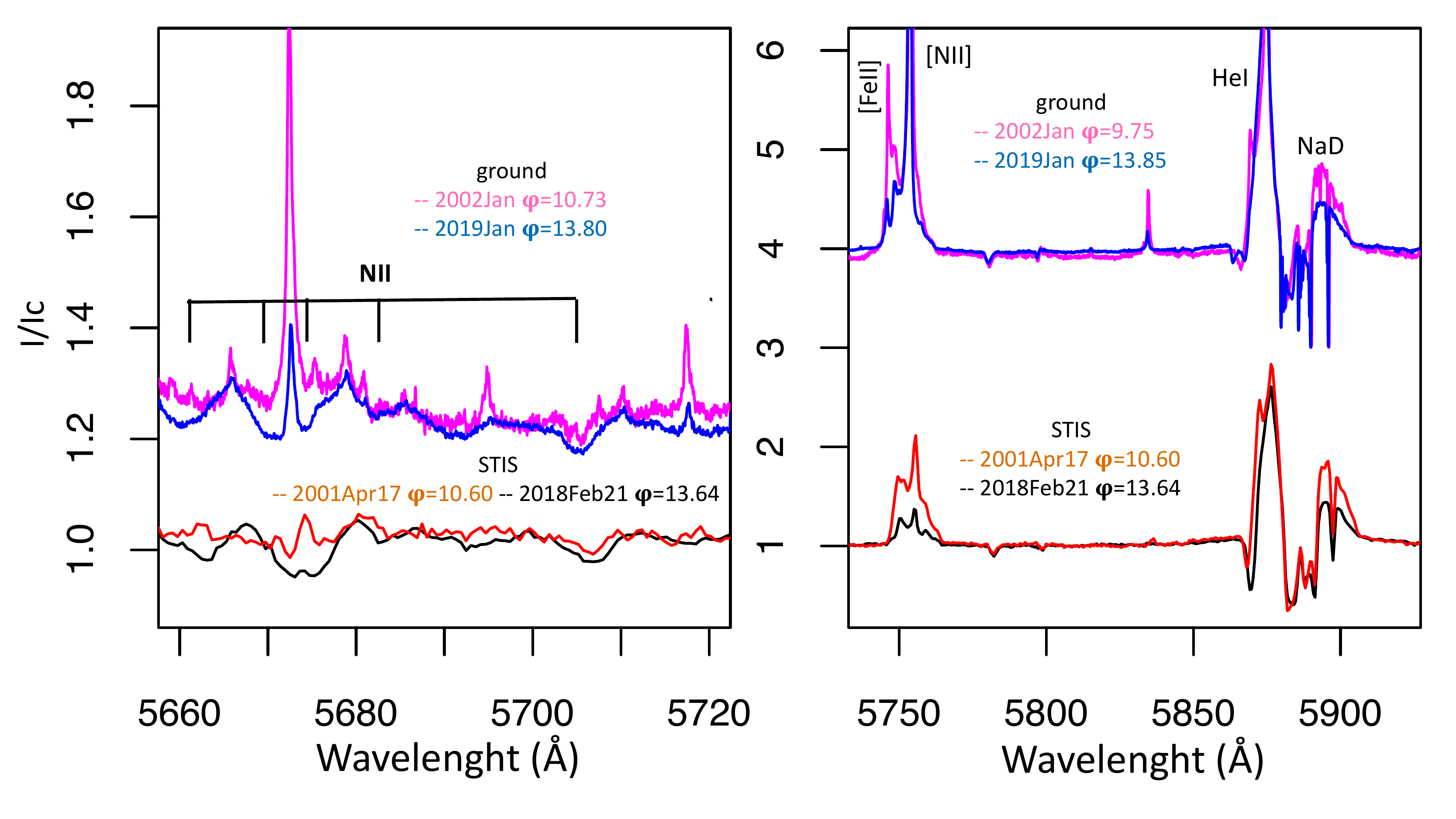}}
\caption{ Comparison of spectral evolution obtained from HST/STIS spectra taken at phases 10.60 and 13.64, and ground-based spectra taken at 10.73 and 13.80. The broad emission components show similar weakening in space and ground-based data, as also is the absorption strengthening in \ion{He}{i}\,$\lambda$5875 (right panel). Narrow emission affects more the ground-based spectra but also fills most of the \ion{N}{ii} absorption P~Cyg components (left panel). \ion{N}{ii}\,$\lambda$5710 is almost free of contamination in space and ground-based spectra, showing similar time-strengthening behavior in both datasets. Extraction in a lower number of rows could lead to less contamination in STIS spectra, but this would have demanded much longer exposures to achieve a high signal-to-noise ratio.}
\label{groundstis}
\end{figure*}

{
Line strengths were measured as follows: $a)$ the line depth intensity was measured directly at the minimum of the P Cygni absorption component relative to the stellar continuum and reduced by subtracting 1 (negative strength); $b)$ the peak of broad emission lines was deblended from the narrow component by a double Gaussian fit (narrow plus broad components) and the peak of the broad component is the peak of the broader Gaussian. This peak also was subtracted by the continuum (set at 1.0). This method was applied to pairs of 1993.5 and 2021.2 ground-based spectra and 2001.4 and 2018.2 STIS spectra.}

\section{Results}
\label{sectionresults}

Previous studies of P~Cygni absorption variations on \ion{He}{i}~$\lambda$6678 line \citep{Mehner2010,mehner12}, \ion{He}{i}~$\lambda$4713 and  \ion{N}{ii}~$\lambda$4601-46743 \citep{mehner15} have been limited to { three} orbital cycles. Here, we examine a larger number of spectral lines with a denser and longer time-sampling to monitor the spectral evolution in the optical window, using ground-based observations.

The absorption components exhibit four types of variability: a) variability in the depth of the P~Cygni absorption components around  periastron; b) variations along the orbital cycle due to changes in the distorted shape of the primary star's wind caused by the WWC cavit -- In most of the lines, these variations have a relatively low amplitude when compared to the low-excitation event around periastron; c) episodic low-amplitude variations near mid-cycle, lasting from days to months; d) continuous, smooth, long-term low-amplitude modulation.

The events caused by periastron passage are easily identifiable by the strong peaks in the EW curves -- the most noticeable variations start ten days before phase zero and last for about three months -- see Fig.~\ref{timeseries}. These events are characterized by an increase in the radial velocities of the absorbing components beyond the terminal wind speed of the primary star \citep[420\,\kms;][]{groh12} - see Fig.~\ref{lineprof}. 

Episodic variations are difficult to identify as they require frequent monitoring of the EW over long time scales (years). They do not show changes in the velocity of the line profiles in absorption at mid-cycle. We identify those sporadic variations by analyzing the time series after subtracting the long-term trend. 

Long-term variations in absorption lines can be gleaned by comparing spectra taken at the same phase in high-excitation phases, in different orbital cycles, as such variations are generally larger than the sporadic low-amplitude variations. The long-term increase in the strength of the P~Cygni absorption component is easy to see in some lines, such as \ion{He}{i}~$\lambda$10830, but more difficult in other lines such as \ion{He}{i}~$\lambda$5875 and \ion{H}{i}~Pa12\,$\lambda$8752.

Our planned modeling of the spectral evolution is based on comparison of line profiles of two spectra taken at the same phase, sufficiently separated in time as to have clear differences. To validate this approach and the selection of the specific dates we follow three steps. First we compare STIS spectra with ground-based spectra to show that they deliver compatible results, at least for the very rare dates when this is feasible. Second, we compare ground-based line profiles of spectra taken at the same phases in different cycles to show that the strengthening of the P~Cygni absorption is a general property of of the spectral evolution. Third, we construct long time series of EWs, to see what phases are most repeatable, except for the long term effect.

\subsection{ Are ground-based spectra comparable to STIS spectra to study long-term spectral evolution? }
\label{groundSTIS}

{ Although the ground-based spectra have much more nebular contamination than the STIS spectra, both data sets display a similar spectral evolution. This can be shown by comparing the evolution in STIS spectra that were taken at the same phase during the high-excitation state, with the evolution in ground-based spectra, shown in Fig.~\ref{groundstis}. For this exercise, we use STIS observations that were taken at $\phi\,=\,10.60$ and 13.64. while the almost coeval ground-based spectra were taken at phases 10.7 and 13.80. The exact amount of line variation is not important here. The right panel of Fig.~\ref{groundstis} shows that the P~Cyg absorption in \ion{He}{i}$\lambda$5875 increases with time in both STIS and ground-based spectra while [\ion{N}{ii}]$\lambda$5755 and \ion{NaD} complex decrease. The left panel shows the faint lines of \ion{N}{ii}. The group member at $\lambda$5712 shows a small degree of increase with time. The other four members also show the same trend of increase but are contaminated by nebular emissions. We tried to diminish the length of the extraction window to 0\farcs1 to improve the situation, slightly reducing the central peak of the $\lambda$5776+5680 blend, but at the expense of 
decreasing the signal-to-noise ratio. This set of \ion{N}{ii} lines was reported by \citet{mehner11}, who also analyzed a STIS spectrum. They are much more evident than in the spectrum we present here. The reader must take in mind that that spectrum was taken only five months after periastron passage, when the absorption lines are stronger, with a much longer time exposure and adopting a smaller( 0\farcs1) extraction window, leaving almost all the nebular contamination outside the slit aperture.

 To confirm the trend of strengthening of the absorption in this line, we call attention to the report made by \citet{davidson15}, who examined another set of \ion{N}{ii} lines in the range $\lambda$4601-4643\,{\AA}. They were not evident in 2003.3 and 2009.0, but appeared as stronger absorption features in 2014.5. Those lines are stronger than in 2014.5, even in ground-based spectra taken during the 2020 periastron. Such a comparison indicates that ground-based spectra show similar trends to those observed from STIS observations, even in a challenging situation like in the P~Cygni absorption of \ion{N}{ii}. This study validates our choice for using the ground-based spectra because they comprise many cycles with a better time sampling of the mid-cycle phases.} 

\subsection{ Comparison between line profiles taken at the same phase in different orbital cycles }
\label{subs-lineprof}

\begin{figure*}[ht!]
\centering
{\includegraphics[width=\linewidth, angle=0,viewport=0bp 45bp 840bp 520bp, clip]{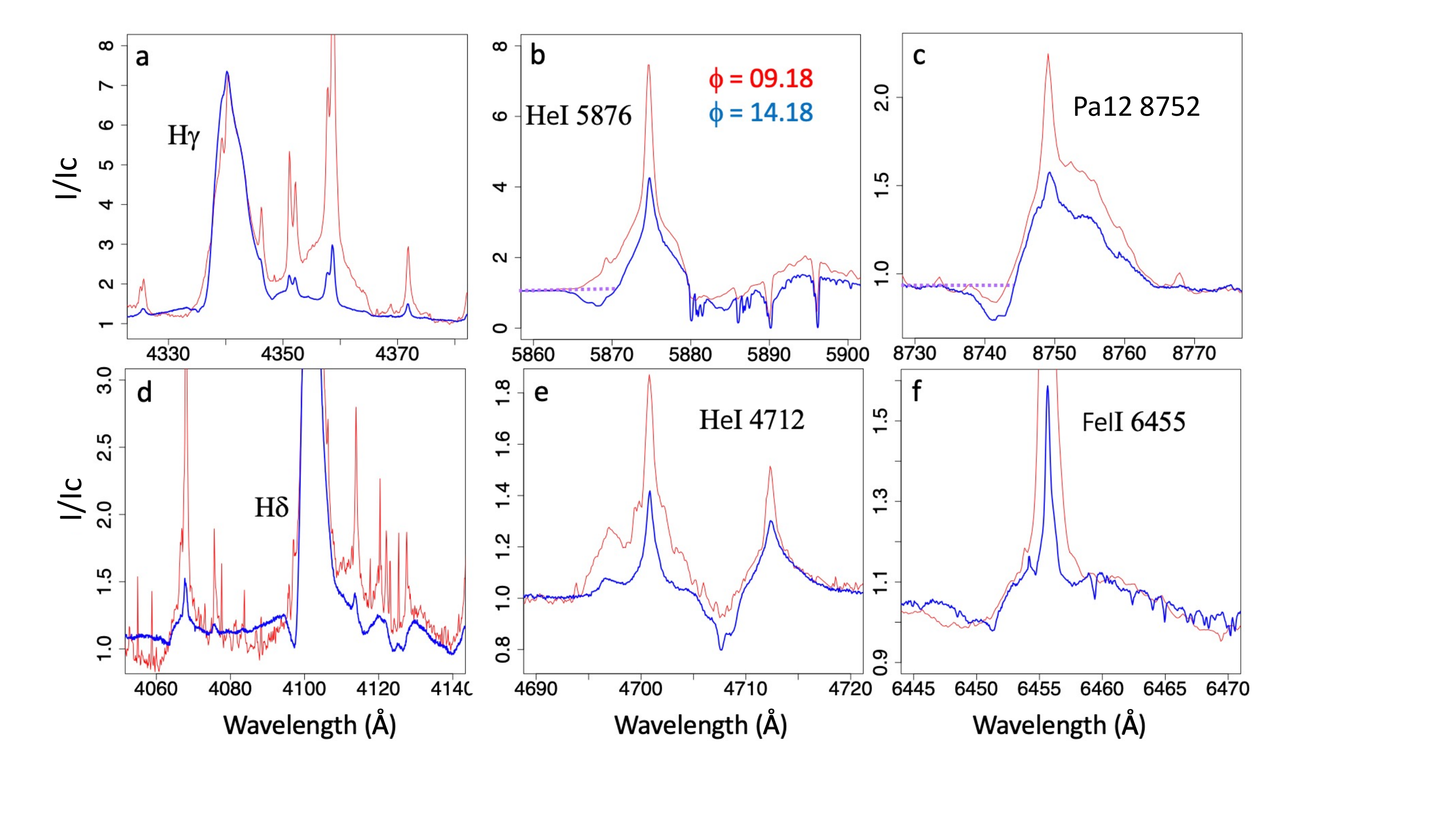}}
\caption{Representative line profiles at $\phi$~=9.18 -- June 1993 (thin red lines) and $\phi$~=14.18 -- March 2021 (thick blue lines), showing the increase in P~Cygni absorption over the 28-year time interval and a decrease of (almost all) emission lines at mid-cycle.  Also evident is the dramatic weakening of the narrow nebular lines, and the broad base associated with forbidden and permitted \ion{Fe}{ii} lines. For flux-calibrated spectra, see Fig. 2 of \citet{damineli21}. Horizontal dotted lines in panels $b$ and $c$ indicate the continuum used for EW absorption measurement.}
\label{5cycles}
\end{figure*}

\begin{figure}[ht!]
\centering
{\includegraphics[height=0.95\linewidth,angle=-90,viewport= 0bp 65bp 840bp 410bp, clip]{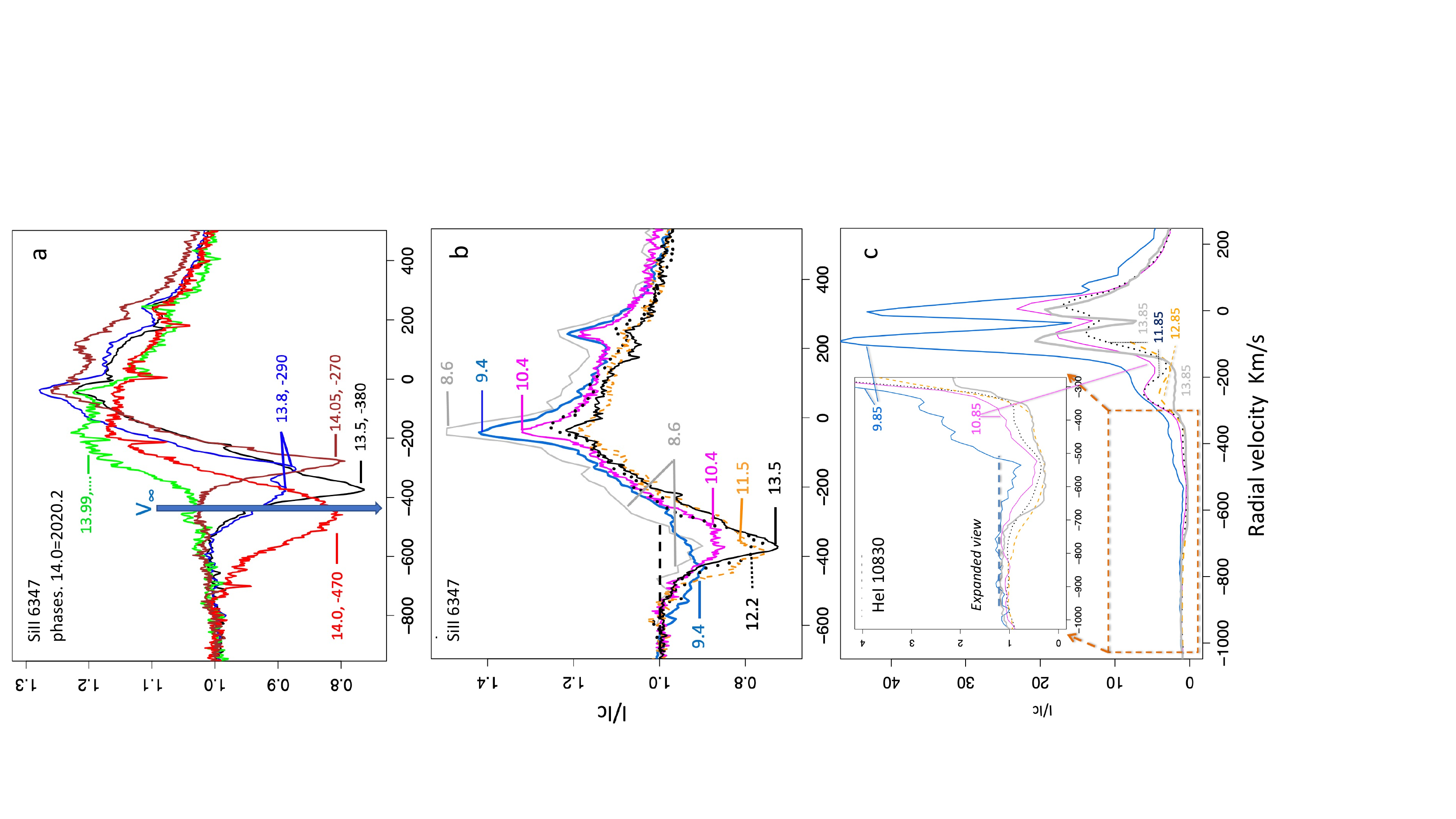}}
\caption{Line profiles of \ion{Si}{ii}~$\lambda$6347 (panels a and b) and of \ion{He}{i}~$\lambda$10830 (panel c). The inset in panel b displays an expanded view of the absorption profile. Panel a shows variability around periastron. Panels b and c show high-excitation phases. The horizontal dotted line in the inset indicates the adopted stellar continuum used for EW absorption measurements.}
\label{lineprof}
\end{figure}

\begin{figure*}[ht!]
\centering
{\includegraphics[width=17.5 cm, angle=0,viewport=20bp 10bp 430bp 525bp, clip]{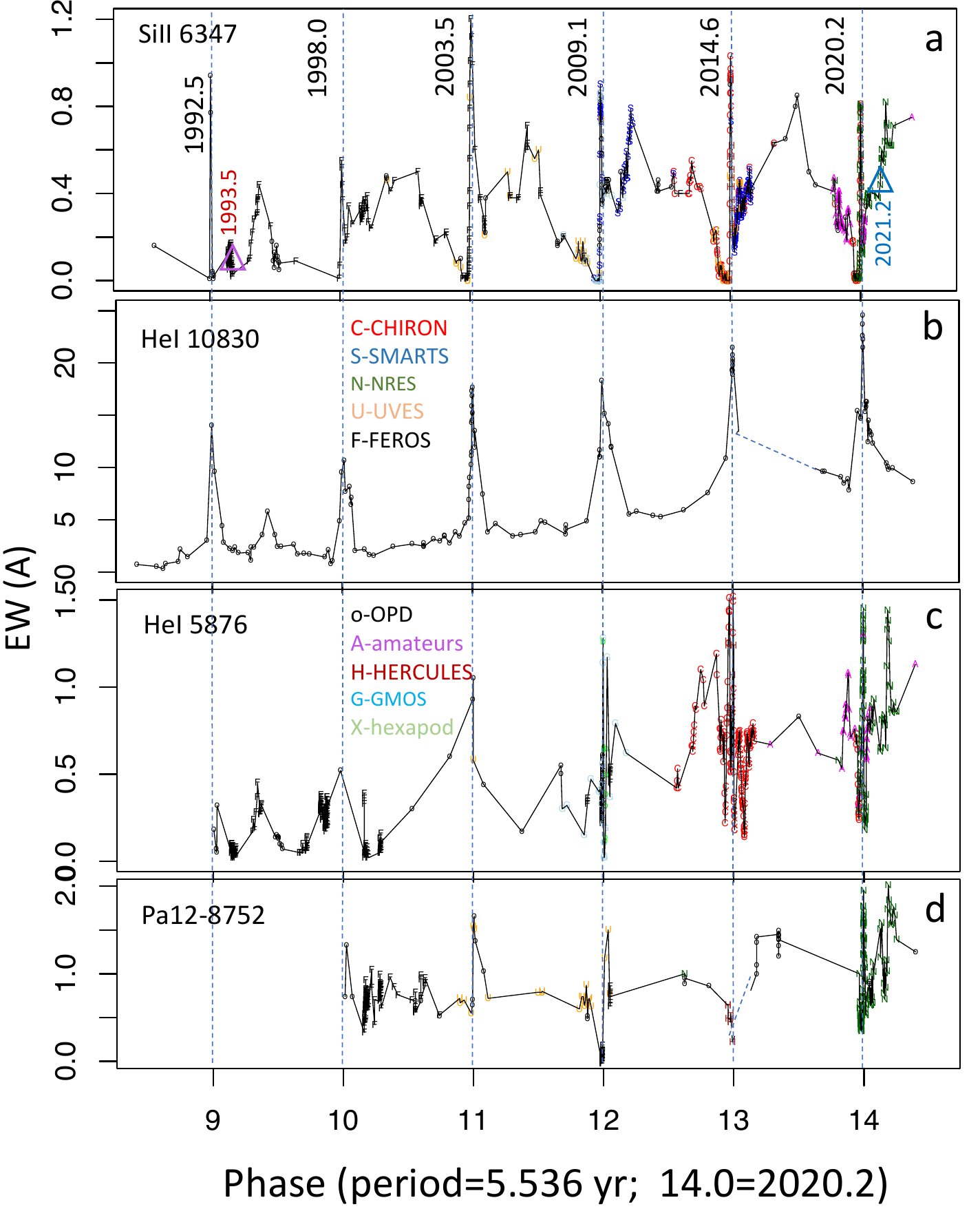}}
\caption{Time series of EWs of the P~Cygni component in absorption. 
a) \ion{Si}{ii}~$\lambda$6347: this line indicates that variability is complex even at high excitation phases; b) \ion{He}{i}~$\lambda$10830: strengthens smoothly at high excitation; c)\ion{He}{i}~$\lambda$5875: although the general trend is of increasing strength, there are occasional oscillations outside of periastron; d) Paschen\,12: shows transient variations in the vicinity of periastron. Observations from different facilities are labeled by letters and colors - a zoomed view shows more clearly the letters.
The two triangles in the upper panel indicate the dates selected for modeling the spectra - which in reality was replaced by the FOS\,4 reflected spectrum which has no narrow lines.}
\label{timeseries}
\end{figure*}

Visual comparison between (a window of) two high-resolution spectra taken 28 years apart (five orbital cycles) are presented at the bottom plot of Fig.~\ref{4500_dirXrefl}. The two ground-based spectrographs, ESO/FEROS (at the 1.5$-$m ESO telescope) and LCOGT/NRES (1-m telescope), have equal spectral resolution R\,$=$\,48,000 and are fiber-fed with projected entrance diameter projected on the sky of $\sim$\,2\farcs7.
They were taken one year after periastron, the first in June 1993  ($\phi$\,$=$\,9.18)\footnote{In this work, we adopted the period $P$\,$=$\,2022.7~d reported by \citet{teodoro16}, $T0$\,$=$\, 2456874.4 ($\pm$ 1.3 days) and the \citet{groh04} orbital cycle numbering scheme, in which cycle \#~14 begins at the last periastron passage on February 18, 2020.} and the second in March 2021 ($\phi\,=\,14.18$). At this phase, the spectrum has already recovered from the ``low-excitation event". The 1993 spectrum is representative of what has been called the ``\ec-like'' spectrum.

{ In the $\phi\,$=$\,9.18$ spectrum, absorption profiles only appeared in lines from very high-excitation levels of the Balmer and Paschen series, in \ion{Si}{ii}\,$\lambda$6347, and in \ion{He}{i} lines such as  \ion{He}{i}\,$\lambda$4026, \ion{He}{i}\,$\lambda$4712 and \ion{He}{i}\,$\lambda$10830. During cycles 9 and 10, strong absorption components appeared only around periastron passage -- see Fig. \ref{lineprof}. }

{ The long-term strengthening of P~Cygni components affects { many other lines. For example,} Fig.~\ref{5cycles} focuses on line-profile differences at $\phi\,=\,9.18$ and $\phi\,=\,14.18$ for some representative lines. Panel $d$ (bottom left panel) shows that the \hd line did not exhibit P~Cygni absorption at $\phi\,=\,9.18$, but did five orbital cycles later for the same phase. The top-right panel of Fig.~\ref{5cycles}\,$c$  shows that the P~Cyg absorption of Pa\,12 line - (\ion{H}{i}~$\lambda$8752) - has strengthened, as was the case for \hd,  \hg, \ion{Fe}{ii}~$\lambda$6455, \ion{He}{i}~$\lambda$5875 and \ion{He}{i}~$\lambda$4712. Panels $b$ and $c$ show a dotted line representing the local stellar continuum used to integrate the EW along the absorption-line profiles.}

Some spectral lines, such as \ion{Si}{ii}~$\lambda$6347 and \ion{He}{i}~$\lambda$10830, were monitored sufficiently often to allow a comparison of their line profiles over five or six orbital cycles. The top panel of Fig.~\ref{lineprof} shows that far from periastron, velocities in the absorption profiles are smaller than the terminal velocity of {\ec} primary's wind, which is compatible with line formation at smaller radii. From 10 days before to a month after periastron, the profile shows very { high} speeds, as expected if the trailing arm of the WWC wall passes through our LOS at those times. The middle panel of Fig.~\ref{lineprof} shows that the P~Cygni component of \ion{Si}{ii}~$\lambda$6347 at phase 0.4\,$-$\,0.6 increases with time. The bottom panel of Fig.~\ref{lineprof} shows the same strengthening long-term evolution of \ion{He}{i}~$\lambda$10830 P~Cygni component, in this case at $\phi$\,=\,0.85. An expanded view of the P~Cygni absorption trend is shown in the inset of panel c. 

The evolution of these two spectral lines shows that the amplitude of both emission and absorption variations seem to have slowed down in the last three cycles, consistent with the almost complete disappearance of many ``circumstellar" absorption lines in the UV \citep{Gull2021,gull23}, and with the disappearance of the blue-shifted narrow absorption of NaD \citep{pickett2022}.

\subsection{ EW time series of P~Cygni absorption components}
\label{subsectionTimeSeries}

The comparison of spectra taken at the same phase in successive cycles is very useful because, in principle, they would be identical if no long-term evolution occurred. However, as we are interested in low-amplitude effects, even minor irregularities in the orbital variations can impact the comparisons. A more panoramic view would be assembling time series in which such irregularities could be identified. The number of spectral lines in which this is possible over five to six cycles is small, especially as only a few lines show an absorption component at mid-cycle. In addition, { since the measurements come from different instruments, we had planned to scale the different series in epochs when they are coeval. However, scaling was not necessary in the case of some lines and not possible for others, as seen ahead. The plots of the EW time series (Fig.~\ref{timeseries}) have the identification of every observational point with a letter indicating its instrumental source - a zoomed view might be necessary to see clearly the individual points. }

{ The EW-curve of \ion{Si}{ii}~$\lambda$6347 shows an overall long-term increase but with plenty of structures repeated over every orbital cycle (panel $a$ of Fig.~\ref{timeseries}). We call attention that the mid-cycle maximum does not occur at the exact same phase in every cycle. This exemplifies how careful the choice of dates to examine long-term evolution must be.} The crossing of the trailing arm of the WWC causes sharp enhancement of \ion{Si}{ii}~$\lambda$6347 absorption around periastron passage. 

The P~Cygni absorption component that has the simplest time variation is that of \ion{He}{i}~$\lambda$10830, as shown in panel b of Fig.~\ref{timeseries}. { The low scatter of this EW-curve is explained by the large intensity of this feature and, in part, because it is recorded by the same spectrograph (OPD/Coud\'e) over three decades, although using several different CCD detectors. The EW-curve shows a smooth long-term increase at mid-cycle phases, combined with the usual narrow peaks at periastron.}

{ The \ion{He}{i}~$\lambda$5875 absorption component strengthens with time, but it exhibits irregularities even at mid-cycle, differently from \ion{He}{i}~$\lambda$10830 (see the middle panel of Fig.~\ref{timeseries}).} This is due to the contribution of the WWC in this spectral line as shown by \citet{richardson16}. These variations are even more pronounced at periastron passages and are not strictly periodic. Even so, it is possible to identify a long-term strengthening trend in the mid-cycle from cycle \#10 onward. The modulus of the radial centroid velocity of the P~Cygni absorption component is $\sim$\,500\,\kms at mid-cycle and up to 1000\,\kms at periastron -- well  beyond the terminal speed of the wind -- indicating that they are probably produced in the WWC walls and are subject to shock instabilities. This finding aligns with the results from \citet{richardson16}.}

{ Even though measurements of weak lines are subject to large uncertainties,} the absorption component of the \ion{H}{i}Pa12$\lambda$8752 line also { strengthens} with time, as shown in the bottom panel of Fig.~\ref{timeseries}. Since this \ion{H}{i} line is formed in the innermost parts of the primary's wind, the emission contribution from radii outside the occulter is smaller than in lower \ion{H}{i} transitions. { Studying this spectral line is important} because it has a permanent {and mostly non-blended} absorption component.

\begin{table}
\centering		
\caption{Sample of Tables A.1 -- A.4 in online material}
\label{onlinetables}
\begin{tabular}{cccc} 
\hline
HJD	 &	EW 5875	&	$\sigma_{\rm{EW}}$	&	Facility	\\
     & {(\AA)}  & {(\AA)}                &           \\
\hline
48793.994	&	0.18	&	0.09	&	OPD/Coud\'e\\
48830.020	&	0.07	&	0.06	&	OPD/Coud\'e	\\
48838.996	&	0.05	&	0.04	&	OPD/Coud\'e	\\
48843.988	&	0.32	&	0.06	&	OPD/Coud\'e	\\
49063.609	&	0.05	&	0.04	&	ESO/FEROS	\\
\hline
\end{tabular}
\begin{tabular}{cccc} 					
HJD	 &	EW 6347	&	$\sigma_{\rm{EW}}$	&	Facility	\\
     & {(\AA)}  &{(\AA)}                &           \\
\hline
47918.500	&	0.16	&	0.03	&	OPD/Coud\'e	\\
48780.500	&	0.01	&	0.00	&	OPD/Coud\'e	\\
48793.990	&	0.94	&	0.19	&	OPD/Coud\'e	\\
48798.500	&	0.77	&	0.15	&	OPD/Coud\'e	\\
48825.010	&	0.04	&	0.01	&	OPD/Coud\'e	\\
\hline
\end{tabular}
\begin{tabular}{cccc} 						
HJD	 &	EW 10830	&	$\sigma_{\rm{EW}}$	&	Facility	\\
     & {(\AA)}  &{(\AA)}                &           \\
\hline
47613.590	&	0.72	&	0.12	&	OPD/Coud\'e	\\
47916.730	&	0.54	&	0.09	&	OPD/Coud\'e	\\
48020.800	&	0.32	&	0.05	&	OPD/Coud\'e	\\
48059.390	&	0.79	&	0.13	&	OPD/Coud\'e	\\
48255.760	&	1.00	&	0.17	&	OPD/Coud\'e	\\
\hline
\end{tabular}
\begin{tabular}{cccc} 						
HJD	 &	EW Pa-12	&	$\sigma_{\rm{EW}}$	&	Facility	\\
 & {(\AA)}  &{(\AA)}                &           \\
\hline
50833.750	&	0.74	&	0.15	&   OPD/Coud\'e	\\
50854.583	&	1.33	&	0.20	&	OPD/Coud\'e	\\
50946.375	&	0.74	&	0.15	&	OPD/Coud\'e	\\
51131.815	&	0.33	&	0.16	&	ESO/FEROS	\\
51135.819	&	0.45	&	0.23	&	ESO/FEROS	\\
\hline
\end{tabular} \\
{{Note.} The first five rows of each table are presented. The full version of the table is available online.}
\end{table}

The EW time series provides a robust method to study the long-term strengthening of P~Cygni absorption features, which was shown in previous reports \citep{Mehner2010,mehner12,mehner15}, but were restricted to shorter time intervals. EW time series are important to track the long-term evolution of the P~Cygni absorption components. { Still, many lines exhibit relatively weak features and/or are severely blended with other spectral components. This is why we rely on} line peak/depth variation for modeling purposes.   

{ The general evolution pattern of the observed spectrum in the last 30 years is $a$) a strong weakening of the narrow emission lines; $b$) a general weakening of the broad emission line components by 50\% for lines formed up to large radii  in the wind (low excitation lines) and less for lines of high excitation transitions, formed close to the central stars; $c$) a general strengthening of P~Cygni absorption components. This can be seen for representative lines in Fig. \ref{obsXmodels} (magenta and green lines) and quantitatively (rows $1$ and $2$) of Table~\ref{table_modelXobs}. In this table, in addition to the line strength (rows 1 to 6), we present the percent difference between spectra (rows 7 to 12), as indicated in column 1. Negative numbers are produced when the line in the first spectrum is weaker than in the second. Numbers printed in red indicate variations in the contrary sense of observations. Differences smaller than 20\% are disregarded.}

\section{\texorpdfstring{{\sc cmfgen}}{CMFGEN} modeling}
\label{section_cmfg}

The main goal of this paper is to show that the dissipating occulter modulates the strength of the emission and P~Cygni absorption strengths in accordance with the observed long-term spectral evolution of \ec. Since only the general trend of the $V$-band extinction is available \citep{damineli19}, only the relative line strength is useful, not a time series with every single EW measurement. After a careful examination of the time series in { Sect.\,\ref{sectionresults}}, we chose two ground-based spectra taken in 1993.5 ($\phi$\,=\,9.18) and in 2021.2 ($\phi$\,=\,14.18) to represent the ratio of the line-strength variation. Other phases in mid-cycle could be used, but 0.18 is the one for which we have the longest time interval - maximizing the spectral evolution - in addition, at this phase -- one year after the periastron -- we look to the primary star photosphere in a less-disturbed situation than in others \citep{madura13}. The WWC trailing arm has already left our LOS, and the cavity is not wholly occupying our view.} Line peaks/depths offer a cleaner way than EWs to measure line strength variations as they are less affected by blend effects on the line wings. 
We call \emph{line strength} the peak (or depth) of the broad wind lines above (or below) the stellar continuum.
These are the quantities we use to compare with the synthetic models and are reported in Table~\ref{table_modelXobs}.

\subsection{\texorpdfstring{{\sc cmfgen}}{CMFGEN} model of the present-day unosbtructed wind}
\label{subsection_fullwind}

We used the {\sc cmfgen} code to calculate a full wind model following \citep{hillier2001, hillier06}, with the following parameters:\\
\indent    R$_\ast$\,$=$\,240\,\rs; \\
\indent   $\dot{M}$\,$=$\,3.9\,$\times$\,10$^{-4}$\,\Msunyr; \\
\indent 
L\,$=$\,4\,$\times$\,10$^{6}$\,\ls; and \\
\indent 
N(H)/N(He)\,$=$\,10\\
\noindent{This model is similar to that which fitted the infrared interferometry across the Brackett alpha emission line (see \citealt{weigelt21} and references therein). The N(H)/N(He) ratio was increased from 5 to 10 to better explain the influence of the occulter on the \ion{Fe}{ii}\ lines. In turn, this required a slight reduction of the mass-loss rate from  5.0\,$\times$\,10$^{-4}$\,\Msunyr to preserve the H/\ion{He}{i}\ spectrum. When a fixed solar mass fraction for iron is adopted, the \ion{Fe}{ii}\ lines are weaker in the model with a high N(H)/N(He) ratio. Qualitatively, the influence of an occulter on a model with N(H)/N(He)\,$=$\,10 or 5 is not that different. Allowing for the radiation field of the binary companion \citep{damineli96,damineli97,damineli08} would provide an alternative means of weakening the strength of \ion{Fe}{ii} emission lines \citep{groh12}.

{
The modeling of the unobstructed spectrum is complicated by several factors, including:}
\begin{enumerate}
\item While we have an estimate of the primary's luminosity (albeit somewhat uncertain due to the unknown efficiency of the conversion of optical/UV light to IR wavelengths and the luminosity of the companion), we do not have a measurement of the intrinsic optical flux emitted by the system at optical and UV wavelengths. 
\begin{figure}[h]
\centering
{\includegraphics[width=13. cm, angle=-90,viewport=20bp 15bp 810bp 1050bp, clip]{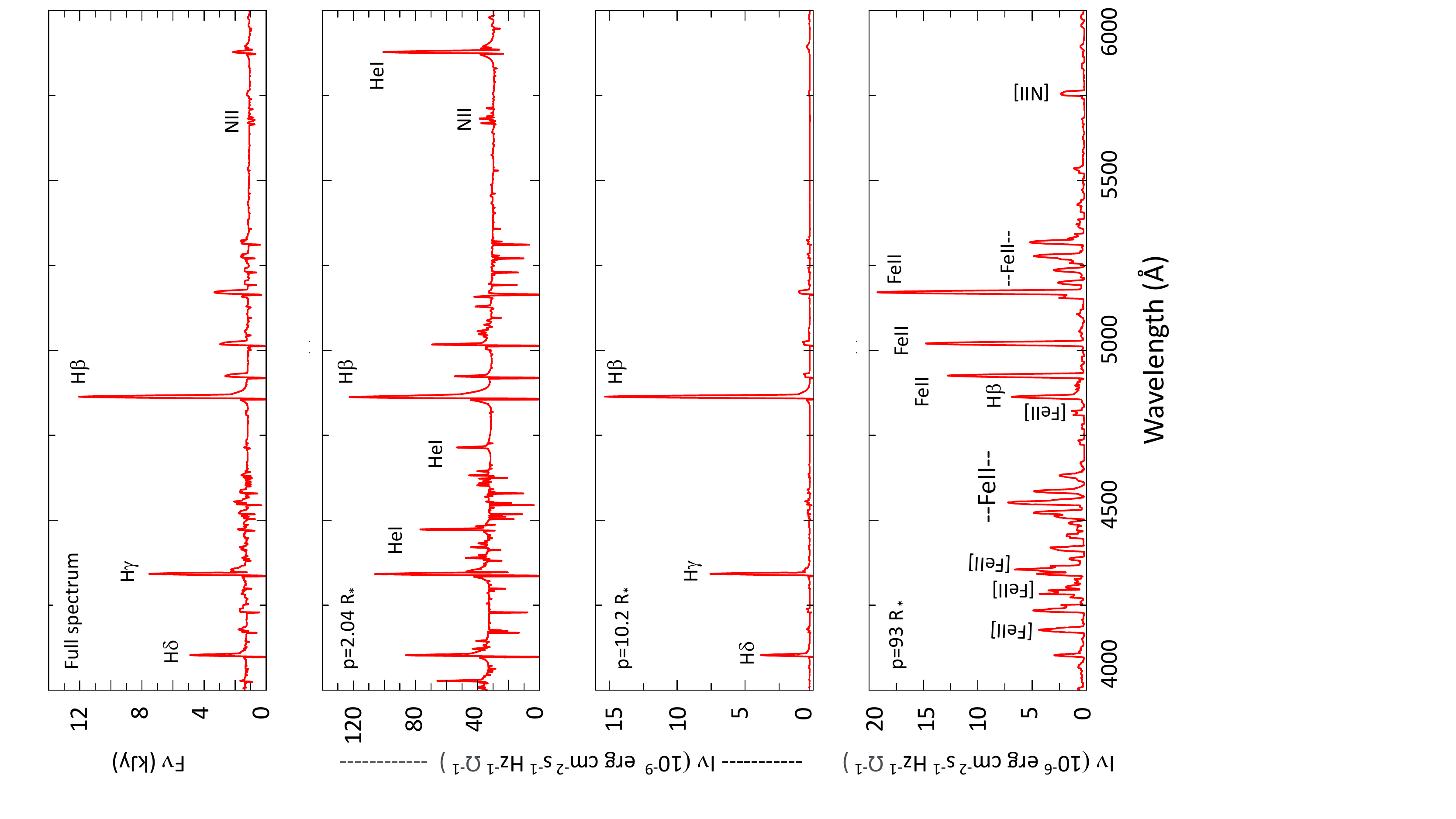}}
\caption{ An illustration of the primary star's fully integrated
spectrum with the specific intensity for different impact parameters, $p$. This parameter is the projected radius inside which the wind is excluded. As can be seen from the figures, the spectra vary strongly with the impact parameter. This variation arises due to radial variations in both ionization and density. In the spectrum
for $p\approx 2 \,R_*$, the \ion{He}{i}\ lines (.e.g., $\lambda 4471$, $\lambda 5876$) line are much stronger than in the stellar spectrum. For  $p\approx 10\,R_*$ H lines dominate the spectrum, while for $p\approx 90\,R_*$,
\ion{Fe}{ii} and [\ion{Fe}{ii}] lines are prevalent. Thus spatial variations in the occulter will have a strong influence on the observed spectrum. Note: If we assume a spherical source, and ignore the occulter, the observed flux is proportional to $\int p I(p) dp$.}
\label{Fig_Ip}
\end{figure}

\item There is a degeneracy between the mass-loss rate and the H/He ratio, such that similar H and \ion{He}{i} spectra can be produced for different mass-loss rates and H/He ratio values \citep{hillier2001}. Metal lines, such as \ion{He}{ii}, can break the degeneracy, but there are additional difficulties in modeling these lines. For example, in models with a higher H/He ratio, the metal lines are weaker since N(H)/N(Z) is higher (the metals are assumed to have a solar mass fraction).
\item The \ion{He}{i} line strengths are most likely affected by the ionizing field of the companion \citep{Nielsen07}, limiting our ability to use the strength of these lines as constraints. Therefore, in the present modeling, we demanded that the \ion{He}{i} lines should be weaker than in the observations.
\item The hotter companion star ionizes the outer wind, affecting the strength of the P~Cygni profiles on Balmer lines, with low-series members (such as \ha) being the most affected. The P~Cyg absorption, especially those associated with \ion{Fe}{ii} lines, is also affected. 
\item The wind is optically thick, so the core radius is challenging to constrain. However, values ranging from 60 to 240~R$_\odot$ are roughly compatible with the observations. In the present modeling, we adopted 240~R$_\odot$ since this allowed us to get a reasonable match to the higher Balmer series members without having too much \ion{He}{i} emission. Note: At a small core radius, the wind dominates the formation of the optical spectrum, but as we increase the radius, the influence of the wind declines, and the shape of the visual continuum changes.
\item If the primary associated with \ec is a fast rotator, there could be significant asymmetries as suggested by \citep[][]{SDG03_lat}. However, direct spectra taken of the central star, and those using the reflected light seen at FOS\,4 are now much more similar than in the past.
\item The occulter may still be influencing (albeit more weakly than in the past) the ground-based spectra. Ground-based spectra at FOS\,4 are also not clean since the spectra did show some evidence of ``scattered'' emission from the Weigelt clumps \citep{hillier92}.
\end{enumerate}

The main effect of the complicating factors is that we cannot develop a well-defined model with realistic error bars. Further, the adopted parameters we selected are generally strongly correlated. However, the parameter space of viable models is limited. Even if { some spectral lines deviate significantly} from observations, this study can produce valid results by looking at the line ratios calculated in different { astrophysical scenarios}: variable mass-loss-rate, presence or absence of a coronagraphic occulter. 

{ The unobstructed (full wind) model is named as f.w.($\dot{M}_{\rm low}$) in row \#3 of Table~\ref{table_modelXobs} and is shown as a black line in Fig.\ref{obsXmodels}. It is intended to fit the present-day spectrum, which we adopted as that observed in reflection at FOS\,4 (green line) - it is very similar to that of 2021.2, except for having less narrow-line contamination and somewhat deeper absorptions. There is a good match to emission lines, except for \ion{He}{i}$\lambda$5875. The broad line component of [\ion{Fe}{ii}]$\lambda$4815 is in reasonable agreement with the observation. The excess emission at the peak of the FOS\,4 spectrum is a reflection from Weigelt clumps. The full wind models produce P~Cygni absorptions stronger than observed. This is a problem not solved in this and previous papers dealing with spectral modeling with {\sc cmfgen}. It may be due to the before-mentioned peculiarities of \ec. }
 
\subsection{How a mass-loss rate that decreases with time contradicts the observed spectral evolution}
\label{subsection_fullwind2}

To explore the effect of a { larger} mass-loss rate in the past, as suggested by other authors (seen\citealt{davidson18} and references therein) trying to explain the decrease in emission line strengths with time, we consider a model with a mass loss rate of  7.8\,$\times$\,10$^{-4}$\,\Msunyr -- twice that of the model that fits the present-day spectrum. All other parameters, including the terminal velocity of the wind, remained fixed. However, LBVs typically show simultaneous changes in both the mass-loss rate and  $V_{\infty}$. No evolution in $V_{\infty}$  has been observed in the spectral evolution of \ec.} 

{ This unobstructed model with a 2$\times$ higher mass-loss rate is hereafter denoted by  f.w.($\dot{M}_{\rm high}$). In Fig.~\ref{obsXmodels} the line profiles of the 2$\times$ mass-loss-rate are represented by the thick grey line, and the percent difference with the model for the present-day spectrum -- f.w.($\dot{M}_{\rm low}$) -- is displayed in the $9^{th}$ row of Table~\ref{table_modelXobs}. }

\begin{figure*}[ht!]
\centering
{\includegraphics[width=1.14\linewidth,angle=-90,viewport=5bp 10bp 840bp 535bp, clip]{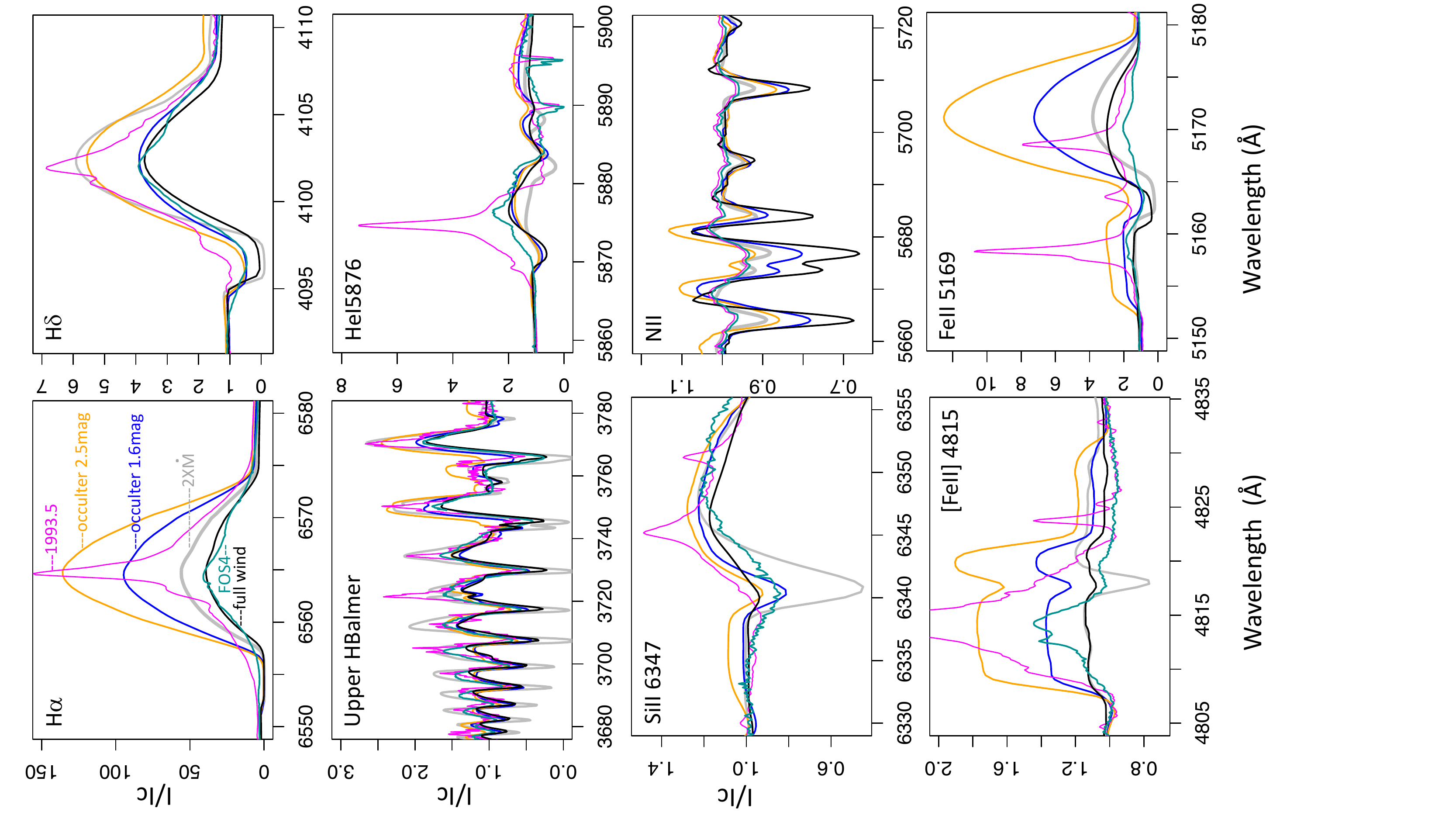}}
\caption{ Spectral comparisons between observations and models. Earlier dates are in magenta and later in green. The \ion{N}{ii} spectrum was taken with STIS and the others are ground-based. The full wind model (black) is intended to be compared with FOS\,4 (unobstructed view) - green - the 2x mass-loss rate (grey) is to be compared with earlier dates; the occulter with 1.6\,mag extinction (blue); and the 2.5\,mag extinction occulter  (orange), both to be compared with earlier dates. }
\label{obsXmodels}
\end{figure*}

{ A decrease in mass-loss rate with time mimics the observed decrease in the observed emission-line components of the Hydrogen Balmer but not in the P~Cygni components, which are weakening in the model, but remained stable in the observations. There is an acceptable agreement between this mass-loss regime and the 1993.5 spectrum for \ion{Fe}{ii}$\lambda$5169, but it predicts a too faint [\ion{Fe}{ii}]$\lambda$4815 emission.
The emission components in the model f.w.($\dot{M}_{\rm high}$) evolve on the contrary sense to the observations for the lines of \ion{He}{i} and \ion{N}{ii}. For the absorption lines of \ion{Si}{ii} and \ion{H}{i}~Pa~12,\ the decreasing mass loss rate also evolves contrary to the observations. Moreover, a decrease in the mass-loss rate is usually accompanied by an increase in $V_{\infty}$, as seen for S~Doradus stars \citep{leitherer1985,groh2009AG,hillier2001}, and it would also change the X-ray light curve contrary to what has been observed in \ec \citep{espinoza22}. 

\subsection{\texorpdfstring{{\sc cmfgen}}{CMFGEN} models with obstruction by the occulter}
\label{subsection_occulter}

{ In this sub-section, we explore the effect of an obscuring "coronagraphic" occulter, which is dissipating or is already dissipated. In general (and assuming the occulter lies outside the spectral formation region) the observed flux is given by
\begin{equation}
    F = \frac{1}{d^2} \int \int_{\scriptstyle \rm star \& wind} I_\nu(x,y) \exp(-\tau_\nu(x,y)) dx dy
\end{equation}
\noindent
where $I(x,y)$ is the specific intensity of the source and $\exp(-\tau_\nu(x,y))$ allows
for the effect of extinction. For a single star $I_\nu(x,y)$ would be azimuthally
symmetric on the sky, but the wind-wind collision zone, and the influence of
the radiation field of the secondary, will break this symmetry. The substantial variation of $I_\nu$ with the impact parameter $p$ is shown in Fig~\ref{Fig_Ip} (which defines the projected internal radius of the excluded wind flux).
Unfortunately, for the choice of $\tau_\nu(x,y)$ we have a wide degree of freedom. The extinction may not be
uniform, and may not be centered on the star. However, for simplicity in the paper we
limit ourselves to uniform occulters that are centered on the star. 

We aim to compare the spectrum of the occulter model with that of the full wind model. We start with the full wind model and subtract from it the spectrum of the region obliterated by a disk. The amount subtracted is defined by the disk size and extinction. We assume that the disk has already dissipated (zero extinction) so that the extinction in 1993.5 is the magnitude difference measured for the entire object (central stellar core plus the Homunculus). The object in 1993.5 was $\Delta$\,V\,$=$\,1.6\,mag fainter than in 2021.2 and hence $A_{V}$\,=\,1.6\,mag \citep{damineli19}. We adjusted the radius of the occulter's disk (r\,$=$\,22.7~R$_{star}$) to produce a good match to the observed \ha line profile in 1993.5. The distance of the occulter from the central star cannot be assessed in this work, so we adopted that suggested by \citet{gull23}: $d$\,$\approx$\,1000~au, formed around 1890 in the same ejection event as the Little Homunculus. 

The occulter models are intended to fit the spectrum taken in 1993.5. Although the occulter1.6 model did a good job for \ha, the upper Balmer emission lines and \ion{He}{i}$\lambda$\,5875 are fainter than observed in 1993.5, and the P~Cygni absorptions are stronger than predicted. Both occulter models are 
 broadly compatible with [\ion{Fe}{ii}]$\lambda$4815 but their predictions are too strong for \ion{Fe}{ii}$\lambda$5169 as compared with observations on 1993.5.

The impact of the occulter on the spectrum cannot be guessed -- it requires a calculation using a realistic model. The complications are many: the occulter is unlikely to be disk causing uniform extinction, and may not be centered on the star. As different transitions originate in different regions of the wind, the impact of the occulter is transition dependent. Further, there is significant free-free emission from the wind which will also be influenced by the
occulter.

\begin{table*}
\centering
\caption{{\bf Line strengths: observations versus models}}
\label{table_modelXobs}
\begin{tabular}{ll|rrrrrrr|rrrrr|} 
    \hline
 & & \multicolumn{7}{c|}{Emission profiles} & \multicolumn{5}{c|}{Absorption profiles} \\
\#&Spectral lines                                               &FeII  &\ha  &\hd   &FeII &[FeII]&HeI  &SiII  &\hd &HeI  &Pa$-$12  &SiII   &NII \\
\hline
& {\bf observations} & \multicolumn{7}{c|}{} & \multicolumn{5}{c|}{} \\
	&		&	8610	&	6563	&	4101	&	4585	&	4815	&	5876	&	6347		&	4101	&	5876	&	8752	&	6347	&	5712	\\
1	&	1993.5	&	4.6	&	61	&	4.5	&	0.8	&	0.8	&	1.9	&	0.36		&	0.00	&	0.00	&	$-$0.07	&	$-$0.05	&	$-$0.04	\\
2	&	FOS4	&	1	&	31	&	2.5	&	0.4	&	0.4	&	1.65	&	0.25		&	$-$0.20	&	$-$0.45	&	$-$0.24	&	$-$0.21	&	$-$0.06	\\
\hline	 		 		 		 		 		 		 		 			 		 		 		 		
& {\bf models} & \multicolumn{7}{c|}{} & \multicolumn{5}{c|}{} \\
3	&	f.w.($\dot{M}_{\rm low}$) 	&	 	&	38	&	2.7	&	0.35	&	0.08	&	0.95	&	0.28		&	$-$0.95	&	$-$0.50	&	$-$0.40	&	$-$0.05	&	$-$0.32	\\
4	&	f.w.($\dot{M}_{\rm high}$) 	&		&	55	&	4.7	&	1.00	&	0.09	&	0.37	&	0.22		&	$-$0.85	&	$-$0.49	&	$-$0.55	&	$-$0.44	&	$-$0.07	\\
5	&	occulter1.6	&		&	66	&	3.2	&	1.25	&	0.23	&	0.86	&	0.21		&	$-$0.90	&	$-$0.50	&	$-$0.46	&	$-$0.16	&	$-$0.17	\\
6	&	occulter2.5	&		&	135	&	4.4	&	3.5	&	0.42	&	0.73	&	0.45		&	$-$0.36	&	$-$0.49	&	$-$0.20	&	$-$0.05	&	$-$0.12	\\
\hline
 {\bf \% diff.} &comparison  & \multicolumn{7}{c|}{} & \multicolumn{5}{c|}{} \\
(2$-$1)/2	&	FOS4:1993.5	&	$-$360	&	$-$97	&	$-$80	&	$-$100	&	$-$100	&	$-$15	&	$-$44		&	100	&	100	&	71	&	76	&	33	\\
(3$-$2)/3	&	f.w.($\dot{M}_{\rm low}$) :FOS4	&&	18	&	$-$11	&	$-$14	&	$-$900	&	$-$74	&	11		&	79	&	$-$28	&	40	&	$-$320	&	81	\\
(3$-$4)/3	&	f.w.($\dot{M}_{\rm low}$):f.w.($\dot{M}_{\rm high}$) 	&&	$-$45	&	$-$74	&	$-$254	&	$-$13	&	{\red 61}	&	{\red 21}		&	11	&	2	&	{\red $-$38}	&	{\red $-$780}	&	78	\\
(3$-$5)/3	&	f.w.($\dot{M}_{\rm low}$):occulter1.6	&&	$-$74	&	$-$19	&	$-$257	&	$-$188	&	9	&	{\red 25}		&	5	&	0	&	$-$15	&{\red $-$220}	&	0	\\
(3$-$6)/3	&	f.w.($\dot{M}_{\rm low}$):occulter2.5	&&	$-$255	&	$-$63	&	$-$900	&	$-$425	&{\red 23}	&	$-$61		&	62	&	6	&	50	&	0	&	63	\\
(6$-$5)/6	&	occulter1.6:occulter2.5	&&$-$640	&	$-$38	&	$-$260	&	$-$83	&	15	&	$-$114&60	&6&65	&	69	&50	\\
\hline
\\
\end{tabular}
{\footnotesize \bf Notes:
a) The line strength is defined by ($I/I_c-1$) at the line peak or deep, negative\,$=$\,absorption, positive\,$=$\,emission;
b)f.w.\,$=$\, full wind; c)$\dot{M}_{\rm low}$\,=\,3.9\,$\times$\,10$^{-4}$\,\Msunyr, $\dot{M}_{\rm high}$\,=\,7.8\,$\times$\,10$^{-4}$\,\Msunyr, occulter~$=$~f.w.($\dot{M}_{\rm low}$) - see the main text; 
d) \% differences: positive\,$=$\,the second is stronger;, negative\,$=$\,the second is weaker). For \ion{N}{ii} we used dates 2001.1 and 2019.1}. Red figures indicate line strength ratios against the observed trend. 
\end{table*}

{ Between 1993.5 and 2021.2 the central star has increased in brightness by $\Delta\,A_V\,$=$\,2.5$\,mag, in order we consider a model where we increase the
 extinction in the core by 2.5\,mag, but keep the same sized occulter. However, this model is unacceptable -- it  delivers a  \ha line intensity that is quite too strong. Keeping the same extinction, we then diminished the size of the coronagraph, obtaining $I/I_c-1\,=\,135$ which is much higher than the observed value 61 in 1993.5, but not absurd.} A disk with r\,=\,10~R$_{star}$ produced a good match to \hd and the upper Balmer series emission, although \ha is still too strong. The \ion{Si}{ii}$\lambda$\,6347 is in reasonable agreement for the emission and absorption component. \ion{N}{ii}$\lambda$\,5712 came better than in occulter1.6 model (occulter's extinction\,$=$\,1.6\,mag), but still too strong. No agreement was found for \ion{He}{i}$\lambda$\,5875 emission. We call this model occulter2.5. As a general comparison, occulter2.5 produces stronger emission and weaker absorption than occulter1.6 (see row 12 in Table~\ref{table_modelXobs} and orange line in Fig.~\ref{obsXmodels}).

Future work could combine these two models with a radial profile of extinction, decreasing from the center to the periphery. However, designing a completely different geometry may be necessary, as \citet{gull23} shows extended patterns of narrow absorption lines in the UV, and \citet{falcke96} reported a knife-edge structure crossing over the central star in the polarization map. 

{ Table~\ref{table_modelXobs} offers a global comparison between observations versus models and between models. Row 7 shows observed variations that are negative for the emission profiles (weakening) and positive for absorptions (strengthening). Rows 9 to 11 show differences in models that should follow the same trend as in row 7, if they do not agree with observational trends, they are printed in red. Variations smaller than 20\% are disregarded. The occulter models with extinction 1.6 disagree with the observational trend in 2 cases and that with 2.5 magnitudes in one case. The 2x mass-loss rate model performed worse than the occulter models to explain the emission line weakening (four cases). Although a complete comparison over the entire spectrum would be more informative (if we had fine-tunned models), this set of lines representative of high and low excitation energies validates the occulter models for explaining the long-term evolution. Although performing worse than the two {\sc cmfgen} occulter models, the decreasing mass-loss rate model cannot be completely ruled out on the basis of spectroscopy alone.}

\section{discussion and conclusions}
\label{sectiondiscussions}

{We have used spectral data taken over 30 years to monitor the long-term evolution of \ec's~spectrum. In particular, we have studied EW time-series of P~Cygni absorption components of \ion{He}{i}~$\lambda$5875 \ion{He}{i}~$\lambda$10830, \ion{Si}{ii}~$\lambda$6347, and \ion{H}{i}~Pa12\,$\lambda$8752.}

Some spectral lines like \ion{He}{i}~$\lambda$5875 and \ion{He}{i}~$\lambda$4712 suffer non-periodic fluctuations \citep{richardson15,mehner15} even outside periastron.  Our ability to distinguish periodic and sporadic variations from long-term variations is enabled by the large amount of data analyzed and because of the significant spectral evolution that has occurred over the last three decades. 

We summarize the observational results from the present work, which support the scenario of a dissipating occulter in front of \ec~A:

\begin{itemize}
 \item Absorption P~Cygni components of a set of representative lines (e.g.: upper members of the Balmer series, \ion{He}{i}~$\lambda$5875, \ion{He}{i}~$\lambda$10830, \ion{Si}{ii}~$\lambda$6347, \ion{N}{ii}~$\lambda$5665-5712, \ion{H}{i}~$\lambda$8752 ) have been strengthening with time.
 
\item P~Cygni absorption components are deeper in reflected than in direct light at high excitation phases, indicating that the stellar wind formed close to the central star suffers additional extinction along our LOS (by the occulter).

\item As the long-term evolution progresses, emission and absorption lines in direct view become more similar to those reflected in the Homunculus (FOS\,4 position), which has decreased much less than in direct light in \ha emission and remained constant for other lines. 

\item  Emission lines that are formed at larger radii from the star (e.g., \ha, \ion{Fe}{ii}) show a { stronger decrease in their} line strength than those formed at smaller radii (e.g., the upper members of the Balmer and Paschen series). Lines formed at very small radii like \ion{N}{ii} and to some extent \ion{He}{i} suffer much smaller line strength changes because the spectral line and its stellar continuum are confined in the same region and change together. 

\item The observed constancy of the central star while the brightness of the Homunculus increased by $\sim$1~mag during the 1940s \citep{Thackeray53,oconnell56}, indicates that the occulter was already in place at that time. The photometric light curve of the stellar core since before that time could be dominated by the occulter's dissipation.

\item {  The rate of decreasing in the emission line intensity and increasing in the absorption maximum/minimum taken at the same phases (see Fig.~\ref{lineprof}) seems to be decreasing slowly with time, indicating that the occulter's dissipation might be already ended or is close to ending. The approaching end of the dissipating process is also hinted by the similarity between the reflected spectrum at FOS\,4 and the present-day direct spectrum as shown in Fig.~\ref{4500_dirXrefl}.}

\item { Although the occulter models examined in this work did not fit all the spectral lines adequately, their percent differences with the full wind evolve in time in the same sense as in observed spectra taken between 1993.5 and 2021.2 in this set of lines (except for \ion{Si}{ii}$\lambda$6347). A refinement of the occulter's structure is likely to give better agreement with observations.}

\item { The decrease in mass-loss rate as the primary explanation for the spectral evolution of \ec is not strongly supported by the {\sc cmfgen} models. It can be rejected because implies in intrinsic changes in the primary star (i.e., changes in terminal wind speed, in the luminosity, in the X-ray light curve, and in the reflected spectrum, at FOS\,4) contrary to observations.} 

\end{itemize}

{ From the Weigelt blob constanc, we know that the occulter provides extinction on scales of less than 0\farcs1 and the occulter should have a smaller radius. Given that H$\alpha$ emission peaks around 0\farcs01, and \ion{Fe}{ii} and  [\ion{Fe}{ii}] emission at even larger radii, it is not unreasonable to assume that differential reddening could influence the EW of absorption and emission lines in the spectrum of \ec, as we have demonstrated in this paper. We speculate that the occulter could be similar to the Weigelt clumps, as it is massive and small and might have formed in the same event. We did not identify narrow emission lines of the occulter, probably because they are at our LOS, differently from the Weigelt clumps which show us their ionized borders. Or it is just an effect of smaller density?}

{ The importance of the occulter has greatly diminished during the lifetime of the HST.
Since 1993. The visual and UV brightness has increased by roughly a factor of 10. The
increased UV brightness has led to an enhanced ionization of the circumstellar material
surrounding \ec, and as a consequence the disappearance of many narrow absorption lines
\citep{gull23}. The increase in visual brightness by a factor of 10 has caused the
EW of emission lines arising in the Weiglet blobs to be reduced by a factor of 10, making them much less visible in the ground-based spectra. Broad forbidden lines of \ion{Fe}{ii}\ have also been greatly reduced in strength. As this brightening has occurred, the direct spectrum has become much more similar to that seen in reflected light \citep{mehner15, damineli21}. It is possible that the occulter has almost dissipated (at least along our LOS). In principle, the reddening to \ec should then approach that of the Weigelt blobs.
 }

 { If the spectral evolution is due to a circumstellar cause, the photometric light curve must be reviewed. \citet{damineli21} claim that the brightening of the star in the last 30 years was due to the decreasing circumstellar extinction.}
 At first glance, the long-term photometric light-curve variability after the 1900s looks contrary to the prediction of a stellar merger (see Fig.\,2 of \citealt{schneider19}) since it predicts a slow decrease of bolometric luminosity as compared to the rapid increase in \ec brightness in the last 30 years. However, suppose these variations are driven by changes in the circumstellar medium, as suggested by the dissipating occulter. One additional test would be the disappearance of HCN molecular absorption in front of the star, reported by \citet{bordiu19} using ALMA.
 
 In that case, the star is much more constant than claimed before and would indicate a dormant LBV, like the other non-S~Dor variable LBVs, including P~Cygni, HD~316285, and HD~168625. In this context, there is no contradiction between the \ec light curve and the binary merger scenario as shown by \citet{schneider19}. In  this scenario, the recovery from the Great Eruption of the 19$^{th}$ century occurred rapidly, on a timescale of less than a century. 

\section{Data Availability}
The complete versions of Tables 1-4 are available in the online journal in a machine-readable format following the CDS/Vizier standards.

\section{Acknowledgments}

We thank D.S.C. Damineli for adapting our plots to color-blind people.
AD thanks to CNPq (301490/2019-8) and FAPESP (2011/51680-6) for their support. The work of FN is supported by NOIRLab, which is managed by the Association of Universities for Research in Astronomy (AURA) under a cooperative agreement with the National Science Foundation.
AFJM is grateful for financial aid from NSERC (Canada). This work is partially based on observations collected with the facilities listed below. DJH gratefully acknowledges support from STScI grants HST-AR-14568.001-A and HST-AR-16131.001-A. CMPR acknowledges support from NATA ATP grant 80NSSC22K0628 and NASA Chandra Theory grant TM2-23003X.

{ Based on observations collected at the European Southern Observatory, Chile under Prog-IDs: UVES:  60.A-9022(A), 70.D-0607(A), 71.D-0168(A), 072.D-0524(A), 074.D-0141(A), 077.D-0618(A), 380.D-0036(A), 381.D-0004(A),282.D-5073(A, B, C, D, E), 089.D-0024(A), 592.D-0047(A, B, C). The first three programmes were described in \citet{stahl05} and the others in \citet{mehner15}; FEROS (partial list): 00.A-0000(A), 69.D-0378(A),, 69.D-0381(A), 71.D-0554(A),079.D-0564(C), 079.A-9201(A), 081.D-2008(A), 082.A-9208(A), 082.A-9209(A), 082.A-9210(A), 083.D-0589(A), 086.D-0997(A), 087.D-0946(A),089.D-0975(A), 098.A-9007(A). Spectra can be downloaded from  the ESO  database by using the instrument name (UVES or FEROS) plus the Julian Date provided in the Appendix Tables.}

Based on observations made at the Coud\'e focus of the 1.6-m telescope for the Observat\'orio do Pico dos Dias/LNA (Brazil).

Based in part on data from Mt. John University Observatory: MJUO -University of Canterbury - New Zealand).

Based in part on observations obtained through NOIR Lab (formerly NOAO) allocations of NOAO-09B-153, NOAO-12A-216, NOAO-12B-194, NOAO-13B-328, NOAO-15A-0109, NOAO-18A-0295, NOAO-19B204, NOIRLab-20A-0054, and NOIRLab-21B-0334. This research has used data from the CTIO/SMARTS 1.5m telescope, which is operated as part of the SMARTS Consortium by RECONS (www.recons.org) members Todd Henry, Hodari James, Wei-Chun Jao, and Leonardo Paredes. At the telescope, observations were carried out by Roberto Aviles and Rodrigo.

Based on observations obtained at the international Gemini Observatory, a program of NSF NOIRLab, which is managed by the Association of Universities for Research in Astronomy (AURA) under a cooperative agreement with the National Science Foundation on behalf of the Gemini Observatory partnership: the National Science Foundation (United States), National Research Council (Canada), Agencia Nacional de Investigaci\'{o}n y Desarrollo (Chile), Ministerio de Ciencia, Tecnolog\'{i}a e Innovaci\'{o}n (Argentina), Minist\'{e}rio da Ci\^{e}ncia, Tecnologia, Inova\c{c}\~{o}es e Comunica\c{c}\~{o}es (Brazil), and Korea Astronomy and Space Science Institute (Republic of Korea).

{ Based on observations with the NASA/ESA Hubble Space Telescope, obtained (from the Data Archive) at the Space Telescope Science Institute, which is operated by the Association of Universities for Research in Astronomy, Inc., under NASA contract NAS 5-26555. These two observations are associated with programmes 8619 and 15067 - PI K. Davidson - link: https://mast.stsci.edu/search/ui/\#/hst\ }

The ESO/UVES and Gemini S/GMOS spectra used in this paper were downloaded from the HST Treasury Project archive { at http://etacar.umn.edu/}.

{ We acknowledge with thanks the variable star observations from https://www.aavso.org/ contributed by observers worldwide and used in this research.}

\vspace{5mm}
\facilities{HST, AAVSO, CTIO, Gemini/South, HST, LCOGT, ESO, MJUO, OPD/LNA, SASER}

\bibliographystyle{aasjournal}

\begin{thebibliography}{}
\expandafter\ifx\csname natexlab\endcsname\relax\def\natexlab#1{#1}\fi
\providecommand{\url}[1]{\href{#1}{#1}}
\providecommand{\dodoi}[1]{doi:~\href{http://doi.org/#1}{\nolinkurl{#1}}}
\providecommand{\doeprint}[1]{\href{http://ascl.net/#1}{\nolinkurl{http://ascl.net/#1}}}
\providecommand{\doarXiv}[1]{\href{https://arxiv.org/abs/#1}{\nolinkurl{https://arxiv.org/abs/#1}}}

\bibitem[{{Bordiu} \& {Rizzo}(2019)}]{bordiu19}
{Bordiu}, C., \& {Rizzo}, J.~R. 2019, \mnras, 490, 1570,
  \dodoi{10.1093/mnras/stz2621}

\bibitem[{{Corcoran} {et~al.}(2017){Corcoran}, {Liburd}, {Morris}, {Russell},
  {Hamaguchi}, {Gull}, {Madura}, {Teodoro}, {Moffat}, {Richardson}, {Hillier},
  {Damineli}, \& {Groh}}]{corcoran17}
{Corcoran}, M.~F., {Liburd}, J., {Morris}, D., {et~al.} 2017, \apj, 838, 45,
  \dodoi{10.3847/1538-4357/aa6347}

\bibitem[{{Damineli}(1996)}]{damineli96}
{Damineli}, A. 1996, \apjl, 460, L49, \dodoi{10.1086/309961}

\bibitem[{{Damineli} {et~al.}(1997){Damineli}, {Conti}, \&
  {Lopes}}]{damineli97}
{Damineli}, A., {Conti}, P.~S., \& {Lopes}, D.~F. 1997, \na, 2, 107,
  \dodoi{10.1016/S1384-1076(97)00008-0}

\bibitem[{{Damineli} {et~al.}(2008){Damineli}, {Hillier}, {Corcoran}, {Stahl},
  {Groh}, {Arias}, {Teodoro}, {Morrell}, {Gamen}, {Gonzalez}, {Leister},
  {Levato}, {Levenhagen}, {Grosso}, {Colombo}, \& {Wallerstein}}]{damineli08}
{Damineli}, A., {Hillier}, D.~J., {Corcoran}, M.~F., {et~al.} 2008, \mnras,
  386, 2330, \dodoi{10.1111/j.1365-2966.2008.13214.x}

\bibitem[{{Damineli} {et~al.}(2019){Damineli}, {Fern{\'a}ndez-Laj{\'u}s},
  {Almeida}, {Corcoran}, {Damineli}, {Gull}, {Hamaguchi}, {Hillier},
  {Jablonski}, {Madura}, {Moffat}, {Navarete}, {Richardson}, {Ruiz}, {Salerno},
  {Scalia}, \& {Weigelt}}]{damineli19}
{Damineli}, A., {Fern{\'a}ndez-Laj{\'u}s}, E., {Almeida}, L.~A., {et~al.} 2019,
  \mnras, 484, 1325, \dodoi{10.1093/mnras/stz067}

\bibitem[{{Damineli} {et~al.}(2021){Damineli}, {Navarete}, {Hillier}, {Moffat},
  {Corcoran}, {Gull}, {Richardson}, {Weigelt}, {Morris}, \&
  {Stevens}}]{damineli21}
{Damineli}, A., {Navarete}, F., {Hillier}, D.~J., {et~al.} 2021, \mnras, 505,
  963, \dodoi{10.1093/mnras/stab1398}

\bibitem[{{Davidson} {et~al.}(1995){Davidson}, {Ebbets}, {Weigelt},
  {Humphreys}, {Hajian}, {Walborn}, \& {Rosa}}]{davidson95}
{Davidson}, K., {Ebbets}, D., {Weigelt}, G., {et~al.} 1995, \aj, 109, 1784,
  \dodoi{10.1086/117408}

\bibitem[{{Davidson} \& {Humphreys}(1997)}]{davidson97}
{Davidson}, K., \& {Humphreys}, R.~M. 1997, \araa, 35, 1,
  \dodoi{10.1146/annurev.astro.35.1.1}

\bibitem[{{Davidson} {et~al.}(2018){Davidson}, {Ishibashi}, {Martin}, \&
  {Humphreys}}]{davidson18}
{Davidson}, K., {Ishibashi}, K., {Martin}, J.~C., \& {Humphreys}, R.~M. 2018,
  \apj, 858, 109, \dodoi{10.3847/1538-4357/aabdef}

\bibitem[{{Davidson} {et~al.}(2015){Davidson}, {Mehner}, {Humphreys}, {Martin},
  \& {Ishibashi}}]{davidson15}
{Davidson}, K., {Mehner}, A., {Humphreys}, R.~M., {Martin}, J.~C., \&
  {Ishibashi}, K. 2015, \apjl, 801, L15, \dodoi{10.1088/2041-8205/801/1/L15}

\bibitem[{Davidson {et~al.}(2005)Davidson, Martin, Humphreys, Ishibashi, Gull,
  Stahl, Weis, Hillier, Damineli, Corcoran, \& Hamann}]{davidson05}
Davidson, K., Martin, J., Humphreys, R.~M., {et~al.} 2005, The Astronomical
  Journal, 129, 900, \dodoi{10.1086/427132}

\bibitem[{{Espinoza Galeas} {et~al.}(2021){Espinoza Galeas}, {Corcoran},
  {Hamaguchi}, \& {Russell}}]{espinoza21}
{Espinoza Galeas}, D., {Corcoran}, M.~F., {Hamaguchi}, K., \& {Russell}, C.
  2021, in American Astronomical Society Meeting Abstracts, Vol.~53, American
  Astronomical Society Meeting Abstracts, 204.10

\bibitem[{{Espinoza-Galeas} {et~al.}(2022){Espinoza-Galeas}, {Corcoran},
  {Hamaguchi}, {Russell}, {Gull}, {Moffat}, {Richardson}, {Weigelt}, {Hillier},
  {Damineli}, {Stevens}, {Madura}, {Gendreau}, {Arzoumanian}, \&
  {Navarete}}]{espinoza22}
{Espinoza-Galeas}, D., {Corcoran}, M.~F., {Hamaguchi}, K., {et~al.} 2022, \apj,
  933, 136, \dodoi{10.3847/1538-4357/ac69ce}

\bibitem[{{Falcke} {et~al.}(1996){Falcke}, {Davidson}, {Hofmann}, \&
  {Weigelt}}]{falcke96}
{Falcke}, H., {Davidson}, K., {Hofmann}, K.~H., \& {Weigelt}, G. 1996, \aap,
  306, L17.
\newblock \doarXiv{astro-ph/9601119}

\bibitem[{{Groh} \& {Damineli}(2004)}]{groh04}
{Groh}, J.~H., \& {Damineli}, A. 2004, Information Bulletin on Variable Stars,
  5492, 1

\bibitem[{{Groh} {et~al.}(2009){Groh}, {Hillier}, {Damineli}, {Whitelock},
  {Marang}, \& {Rossi}}]{groh2009AG}
{Groh}, J.~H., {Hillier}, D.~J., {Damineli}, A., {et~al.} 2009, \apj, 698,
  1698, \dodoi{10.1088/0004-637X/698/2/1698}

\bibitem[{{Groh} {et~al.}(2012{\natexlab{a}}){Groh}, {Hillier}, {Madura}, \&
  {Weigelt}}]{groh12}
{Groh}, J.~H., {Hillier}, D.~J., {Madura}, T.~I., \& {Weigelt}, G.
  2012{\natexlab{a}}, \mnras, 423, 1623,
  \dodoi{10.1111/j.1365-2966.2012.20984.x}

\bibitem[{{Groh} {et~al.}(2012{\natexlab{b}}){Groh}, {Madura}, {Hillier},
  {Kruip}, \& {Weigelt}}]{groh12b}
{Groh}, J.~H., {Madura}, T.~I., {Hillier}, D.~J., {Kruip}, C.~J.~H., \&
  {Weigelt}, G. 2012{\natexlab{b}}, \apjl, 759, L2,
  \dodoi{10.1088/2041-8205/759/1/L2}

\bibitem[{{Gull} {et~al.}(2009){Gull}, {Nielsen}, {Corcoran}, {Madura},
  {Owocki}, {Russell}, {Hillier}, {Hamaguchi}, {Kober}, {Weis}, {Stahl}, \&
  {Okazaki}}]{gull09}
{Gull}, T.~R., {Nielsen}, K.~E., {Corcoran}, M.~F., {et~al.} 2009, \mnras, 396,
  1308, \dodoi{10.1111/j.1365-2966.2009.14854.x}

\bibitem[{{Gull} {et~al.}(2021){Gull}, {Navarete}, {Corcoran}, {Damineli},
  {Espinoza}, {Hamaguchi}, {Hartman}, {Hillier}, {Madura}, {Moffat}, {Morris},
  {Nielsen}, {Pittard}, {Pollock}, {Richardson}, {Russell}, {Stevens}, \&
  {Weigelt}}]{Gull2021}
{Gull}, T.~R., {Navarete}, F., {Corcoran}, M.~F., {et~al.} 2021, \apj, 923,
  102, \dodoi{10.3847/1538-4357/ac22a6}

\bibitem[{{Gull} {et~al.}(2023){Gull}, {Hartman}, {Teodoro}, {Hillier},
  {Corcoran}, {Damineli}, {Hamaguchi}, {Madura}, {Moffat}, {Morris},
  {Richardson}, {Stevens}, \& {Weigelt}}]{gull23}
{Gull}, T.~R., {Hartman}, H., {Teodoro}, M., {et~al.} 2023, arXiv e-prints,
  arXiv:2305.13216, \dodoi{10.48550/arXiv.2305.13216}

\bibitem[{{Hillier} \& {Allen}(1992)}]{hillier92}
{Hillier}, D.~J., \& {Allen}, D.~A. 1992, \aap, 262, 153

\bibitem[{{Hillier} {et~al.}(2001{\natexlab{a}}){Hillier}, {Davidson},
  {Ishibashi}, \& {Gull}}]{hillier2001}
{Hillier}, D.~J., {Davidson}, K., {Ishibashi}, K., \& {Gull}, T.
  2001{\natexlab{a}}, \apj, 553, 837, \dodoi{10.1086/320948}

\bibitem[{{Hillier} {et~al.}(2001{\natexlab{b}}){Hillier}, {Davidson},
  {Ishibashi}, \& {Gull}}]{HDI01_eta_wind}
{Hillier}, D.~J., {Davidson}, K., {Ishibashi}, K., \& {Gull}, T.
  2001{\natexlab{b}}, in Astronomical Society of the Pacific Conference Series,
  Vol. 242, Eta Carinae and Other Mysterious Stars: The Hidden Opportunities of
  Emission Spectroscopy, ed. T.~R. {Gull}, S.~{Johannson}, \& K.~{Davidson},
  15--27

\bibitem[{{Hillier} {et~al.}(2006){Hillier}, {Gull}, {Nielsen}, {Sonneborn},
  {Iping}, {Smith}, {Corcoran}, {Damineli}, {Hamann}, {Martin}, \&
  {Weis}}]{hillier06}
{Hillier}, D.~J., {Gull}, T., {Nielsen}, K., {et~al.} 2006, \apj, 642, 1098,
  \dodoi{10.1086/501225}

\bibitem[{{Humphreys} \& {Davidson}(1994)}]{humphreys94}
{Humphreys}, R.~M., \& {Davidson}, K. 1994, \pasp, 106, 1025,
  \dodoi{10.1086/133478}

\bibitem[{{Leitherer} {et~al.}(1985){Leitherer}, {Appenzeller}, {Klare},
  {Lamers}, {Stahl}, {Waters}, \& {Wolf}}]{leitherer1985}
{Leitherer}, C., {Appenzeller}, I., {Klare}, G., {et~al.} 1985, \aap, 153, 168

\bibitem[{{Madura} {et~al.}(2013){Madura}, {Gull}, {Okazaki}, {Russell},
  {Owocki}, {Groh}, {Corcoran}, {Hamaguchi}, \& {Teodoro}}]{madura13}
{Madura}, T.~I., {Gull}, T.~R., {Okazaki}, A.~T., {et~al.} 2013, \mnras, 436,
  3820, \dodoi{10.1093/mnras/stt1871}

\bibitem[{{Martin} {et~al.}(2021){Martin}, {Davidson}, {Humphreys}, \&
  {Ishibashi}}]{martin21}
{Martin}, J.~C., {Davidson}, K., {Humphreys}, R.~M., \& {Ishibashi}, K. 2021,
  Research Notes of the American Astronomical Society, 5, 197,
  \dodoi{10.3847/2515-5172/ac1fff}

\bibitem[{{Mehner} {et~al.}(2012){Mehner}, {Davidson}, {Humphreys},
  {Ishibashi}, {Martin}, {Ruiz}, \& {Walter}}]{mehner12}
{Mehner}, A., {Davidson}, K., {Humphreys}, R.~M., {et~al.} 2012, \apj, 751, 73,
  \dodoi{10.1088/0004-637X/751/1/73}

\bibitem[{{Mehner} {et~al.}(2011){Mehner}, {Davidson}, {Martin}, {Humphreys},
  {Ishibashi}, \& {Ferland}}]{mehner11}
{Mehner}, A., {Davidson}, K., {Martin}, J.~C., {et~al.} 2011, \apj, 740, 80,
  \dodoi{10.1088/0004-637X/740/2/80}

\bibitem[{{Mehner} {et~al.}(2010){Mehner}, {Davidson}, {Humphreys}, {Martin},
  {Ishibashi}, {Ferland}, \& {Walborn}}]{Mehner2010}
{Mehner}, A., {Davidson}, K.~K., {Humphreys}, R.~M., {et~al.} 2010, The
  Astrophysical Journal, 717, L22, \dodoi{10.1088/2041-8205/717/1/l22}

\bibitem[{{Mehner} {et~al.}(2014){Mehner}, {Ishibashi}, {Whitelock},
  {Nagayama}, {Feast}, {van Wyk}, \& {de Wit}}]{Mehner2014}
{Mehner}, A., {Ishibashi}, K., {Whitelock}, P., {et~al.} 2014, A\&A, 564, A14,
  \dodoi{10.1051/0004-6361/201322729}

\bibitem[{{Mehner} {et~al.}(2015){Mehner}, {Davidson}, {Humphreys}, {Walter},
  {Baade}, {de Wit}, {Martin}, {Ishibashi}, {Rivinius}, {Martayan}, {Ruiz}, \&
  {Weis}}]{mehner15}
{Mehner}, A., {Davidson}, K., {Humphreys}, R.~M., {et~al.} 2015, \aap, 578,
  A122, \dodoi{10.1051/0004-6361/201425522}

\bibitem[{{Mehner} {et~al.}(2019){Mehner}, {de Wit}, {Asmus}, {Morris},
  {Agliozzo}, {Barlow}, {Gull}, {Hillier}, \& {Weigelt}}]{mehner19}
{Mehner}, A., {de Wit}, W.~J., {Asmus}, D., {et~al.} 2019, \aap, 630, L6,
  \dodoi{10.1051/0004-6361/201936277}

\bibitem[{{Nielsen} {et~al.}(2007){Nielsen}, {Corcoran}, {Gull}, {Hillier},
  {Hamaguchi}, {Ivarsson}, \& {Lindler}}]{Nielsen07}
{Nielsen}, K.~E., {Corcoran}, M.~F., {Gull}, T.~R., {et~al.} 2007, \apj, 660,
  669, \dodoi{10.1086/513006}

\bibitem[{O'Connell \& S.J.(1956)}]{oconnell56}
O'Connell, D., \& S.J. 1956, Vistas in Astronomy, 2, 1165,
  \dodoi{https://doi.org/10.1016/0083-6656(56)90047-2}

\bibitem[{{Pickett} {et~al.}(2022){Pickett}, {Richardson}, {Gull}, {Hillier},
  {Hartman}, {Ibrahim}, {Lane}, {Strawn}, {Damineli}, {Moffat}, {Navarete}, \&
  {Weigelt}}]{pickett2022}
{Pickett}, C.~S., {Richardson}, N.~D., {Gull}, T.~R., {et~al.} 2022, \apj, 937,
  85, \dodoi{10.3847/1538-4357/ac898f}

\bibitem[{{Richardson} {et~al.}(2015){Richardson}, {Gies}, {Gull}, {Moffat}, \&
  {St-Jean}}]{richardson15}
{Richardson}, N.~D., {Gies}, D.~R., {Gull}, T.~R., {Moffat}, A.~F.~J., \&
  {St-Jean}, L. 2015, \aj, 150, 109, \dodoi{10.1088/0004-6256/150/4/109}

\bibitem[{Richardson {et~al.}(2016)Richardson, Madura, St-Jean, Moffat, Gull,
  Russell, Damineli, Teodoro, Corcoran, Walter, Clementel, Groh, Hamaguchi, \&
  Hillier}]{richardson16}
Richardson, N.~D., Madura, T.~I., St-Jean, L., {et~al.} 2016, Monthly Notices
  of the Royal Astronomical Society, 461, 2540, \dodoi{10.1093/mnras/stw1415}

\bibitem[{{Schneider} {et~al.}(2019){Schneider}, {Ohlmann}, {Podsiadlowski},
  {R{\"o}pke}, {Balbus}, {Pakmor}, \& {Springel}}]{schneider19}
{Schneider}, F. R.~N., {Ohlmann}, S.~T., {Podsiadlowski}, P., {et~al.} 2019,
  \nat, 574, 211, \dodoi{10.1038/s41586-019-1621-5}

\bibitem[{{Smith} {et~al.}(2003{\natexlab{a}}){Smith}, {Davidson}, {Gull},
  {Ishibashi}, \& {Hillier}}]{SDG03_lat}
{Smith}, N., {Davidson}, K., {Gull}, T.~R., {Ishibashi}, K., \& {Hillier},
  D.~J. 2003{\natexlab{a}}, \apj, 586, 432, \dodoi{10.1086/367641}

\bibitem[{{Smith} {et~al.}(2003{\natexlab{b}}){Smith}, {Gehrz}, {Hinz},
  {Hoffmann}, {Hora}, {Mamajek}, \& {Meyer}}]{smith03}
{Smith}, N., {Gehrz}, R.~D., {Hinz}, P.~M., {et~al.} 2003{\natexlab{b}}, \aj,
  125, 1458, \dodoi{10.1086/346278}

\bibitem[{{Stahl} {et~al.}(2005){Stahl}, {Weis}, {Bomans}, {Davidson}, {Gull},
  \& {Humphreys}}]{stahl05}
{Stahl}, O., {Weis}, K., {Bomans}, D.~J., {et~al.} 2005, \aap, 435, 303,
  \dodoi{10.1051/0004-6361:20042547}

\bibitem[{{Teodoro} {et~al.}(2016){Teodoro}, {Damineli}, {Heathcote},
  {Richardson}, {Moffat}, {St-Jean}, {Russell}, {Gull}, {Madura}, {Pollard},
  {Walter}, {Coimbra}, {Prates}, {Fern{\'a}ndez-Laj{\'u}s}, {Gamen}, {Hickel},
  {Henrique}, {Navarete}, {Andrade}, {Jablonski}, {Luckas}, {Locke}, {Powles},
  {Bohlsen}, {Chini}, {Corcoran}, {Hamaguchi}, {Groh}, {Hillier}, \&
  {Weigelt}}]{teodoro16}
{Teodoro}, M., {Damineli}, A., {Heathcote}, B., {et~al.} 2016, \apj, 819, 131,
  \dodoi{10.3847/0004-637X/819/2/131}

\bibitem[{Thackeray(1953)}]{Thackeray53}
Thackeray, A.~D. 1953, Monthly Notices of the Royal Astronomical Society, 113,
  237, \dodoi{10.1093/mnras/113.2.237}

\bibitem[{Vollmann \& Eversberg(2006)}]{vollmann06}
Vollmann, K., \& Eversberg, T. 2006, Astronomische Nachrichten, 327, 862,
  \dodoi{https://doi.org/10.1002/asna.200610645}

\bibitem[{{Weigelt} \& {Ebersberger}(1986)}]{weigelt86}
{Weigelt}, G., \& {Ebersberger}, J. 1986, \aap, 163, L5

\bibitem[{{Weigelt} {et~al.}(2021){Weigelt}, {Hofmann}, {Schertl}, {Lopez},
  {Petrov}, {Lagarde}, {Berio}, {Jaffe}, {Henning}, {Millour}, {Meilland},
  {Allouche}, {Robbe-Dubois}, {Matter}, {Cruzal{\`e}bes}, {Hillier}, {Russell},
  {Madura}, {Gull}, {Corcoran}, {Damineli}, {Moffat}, {Morris}, {Richardson},
  {Paladini}, {Sch{\"o}ller}, {M{\'e}rand}, {Glindemann}, {Beckmann},
  {Heininger}, {Bettonvil}, {Zins}, {Woillez}, {Bristow}, {Sanchez-Bermudez},
  {Ohnaka}, {Kraus}, {Mehner}, {Wittkowski}, {Hummel}, {Stee}, {Vakili},
  {Hartman}, {Navarete}, {Hamaguchi}, {Espinoza-Galeas}, {Stevens}, {van
  Boekel}, {Wolf}, {Hogerheijde}, {Dominik}, {Augereau}, {Pantin}, {Waters},
  {Meisenheimer}, {Varga}, {Klarmann}, {G{\'a}mez Rosas}, {Burtscher},
  {Leftley}, {Isbell}, {Hocd{\'e}}, {Yoffe}, {Kokoulina}, {Hron}, {Groh},
  {Kreplin}, {Rivinius}, {de Wit}, {Danchi}, {Domiciano de Souza}, {Drevon},
  {Labadie}, {Connot}, {Nu{\ss}baum}, {Lehmitz}, {Antonelli}, {Graser}, \&
  {Leinert}}]{weigelt21}
{Weigelt}, G., {Hofmann}, K.~H., {Schertl}, D., {et~al.} 2021, \aap, 652, A140,
  \dodoi{10.1051/0004-6361/202141240}

\end{thebibliography}

\appendix

\section{Extended data tables (online material)}
\label{appendix_tables}

\startlongtable 
\begin{deluxetable}{cccc}
\tablecaption{Equivalent widths of the 5875\,{\AA} absorption feature.}
\label{table-ew1}
\tablenum{A.1}
\tablehead{\colhead{HJD} & \colhead{EW 5875} & \colhead{$\sigma_{\rm{EW}}$} & \colhead{Facility } \\ 
\colhead{} & \colhead{((\AA))} & \colhead{((\AA))} & \colhead{} } 
\startdata
48793.994	&	0.18	&	0.09	&	OPD/Coud\'e\\
48830.020	&	0.07	&	0.06	&	OPD/Coud\'e	\\
48838.996	&	0.05	&	0.04	&	OPD/Coud\'e	\\
48843.988	&	0.32	&	0.06	&	OPD/Coud\'e	\\
49063.609	&	0.05	&	0.04	&	ESO/FEROS	\\
49063.609	&	0.10	&	0.08	&	ESO/FEROS	\\
49065.609	&	0.10	&	0.08	&	ESO/FEROS	\\
49066.621	&	0.10	&	0.08	&	ESO/FEROS	\\
\enddata
\end{deluxetable} 
\startlongtable 
\begin{deluxetable}{cccc}
\tablecaption{Equivalent widths of the 6347\,{\AA} absorption feature.}
\label{table-ew2}
\tablenum{A.2}
\tablehead{\colhead{HJD} & \colhead{EW 6347} & \colhead{$\sigma_{\rm{EW}}$} & \colhead{Facility } \\ 
\colhead{} & \colhead{((\AA))} & \colhead{((\AA))} & \colhead{} } 
\startdata
47918.500	&	0.16	&	0.03	&	OPD/Coud\'e	\\
48780.500	&	0.01	&	0.00	&	OPD/Coud\'e	\\
48793.990	&	0.94	&	0.19	&	OPD/Coud\'e	\\
48798.500	&	0.77	&	0.15	&	OPD/Coud\'e	\\
48825.010	&	0.04	&	0.01	&	OPD/Coud\'e	\\
48830.010	&	0.03	&	0.01	&	OPD/Coud\'e	\\
48838.300	&	0.01	&	0.00	&	OPD/Coud\'e	\\
48843.990	&	0.02	&	0.00	&	OPD/Coud\'e	\\
\enddata
\end{deluxetable} 
\startlongtable 
\begin{deluxetable}{cccc}
\tablecaption{Equivalent widths of the 10830\,{\AA} absorption feature.}
\label{table-ew3}
\tablenum{A.3}
\tablehead{\colhead{HJD} & \colhead{EW 10830} & \colhead{$\sigma_{\rm{EW}}$} & \colhead{Facility } \\ 
\colhead{} & \colhead{((\AA))} & \colhead{((\AA))} & \colhead{} } 
\startdata
47613.590	&	0.72	&	0.12	&	OPD/Coud\'e	\\
47916.730	&	0.54	&	0.09	&	OPD/Coud\'e	\\
48020.800	&	0.32	&	0.05	&	OPD/Coud\'e	\\
48059.390	&	0.79	&	0.13	&	OPD/Coud\'e	\\
48255.760	&	1.00	&	0.17	&	OPD/Coud\'e	\\
48285.670	&	2.15	&	0.24	&	OPD/Coud\'e	\\
48402.410	&	1.46	&	0.25	&	OPD/Coud\'e	\\
48701.760	&	3.00	&	0.33	&	OPD/Coud\'e	\\
\enddata
\end{deluxetable} 
\startlongtable 
\begin{deluxetable}{cccc}
\tablecaption{Equivalent widths of the Pa-12 8750\,{\AA} absorption feature.}
\label{table-ew4}
\tablenum{A.4}
\tablehead{\colhead{HJD} & \colhead{EW Pa-12} & \colhead{$\sigma_{\rm{EW}}$} & \colhead{Facility } \\ 
\colhead{} & \colhead{((\AA))} & \colhead{((\AA))} & \colhead{} } 
\startdata
50833.750	&	0.74	&	0.15	&   OPD/Coud\'e	\\
50854.583	&	1.33	&	0.20	&	OPD/Coud\'e	\\
50946.375	&	0.74	&	0.15	&	OPD/Coud\'e	\\
51131.815	&	0.33	&	0.16	&	ESO/FEROS	\\
51135.819	&	0.45	&	0.23	&	ESO/FEROS	\\
51135.819	&	0.54	&	0.27	&	ESO/FEROS	\\
51135.822	&	0.41	&	0.21	&	ESO/FEROS	\\
51135.825	&	0.50	&	0.25	&	ESO/FEROS	\\
\enddata
\end{deluxetable}

\label{lastpage}
\end{document}